\documentclass[final,5p,times]{elsarticle}
\usepackage{lineno}
\usepackage[bookmarks=false]{hyperref}
\modulolinenumbers[50]
\journal{The Journal of Systems and Software}
\usepackage[linesnumbered,noline,noend]{algorithm2e}
\SetNlSty{}{}{}
\let\oldnl\nl
\newcommand\nonl{%
  \renewcommand{\nl}{\let\nl\oldnl}}
\usepackage{amsmath}
\usepackage{graphicx}
\usepackage{todonotes}
\usepackage{threeparttable}
\newboolean{showcomments}
\setboolean{showcomments}{true}
\ifthenelse{\boolean{showcomments}}
{ \newcommand{\mynote}[2]{
    \fbox{\bfseries\sffamily\scriptsize#1}
    {\small$\blacktriangleright$\textsf{\emph{#2}}$\blacktriangleleft$}}}
{ \newcommand{\mynote}[2]{}}

\newcommand{\tool}[1]{Sip4j}

\usepackage{listings}
\usepackage{soul} 
\definecolor{KWColor}{rgb}{0.37,0.08,0.25}
\definecolor{CommentColor}{rgb}{0.133,0.545,0.133}
\definecolor{StringColor}{rgb}{0,0.126,0.941}
\usepackage{algcompatible}
\usepackage{amsmath}
\usepackage{rotating}
\usepackage{booktabs}
\usepackage{caption}
\usepackage{mathtools}
\usepackage{lscape}
\usepackage{tablefootnote}
\usepackage{subcaption}
\SetKwComment{Comment}{$\triangleright$\ }{}
\SetKwComment{tcp}{$\triangleright$ }{}%

\SetCommentSty{mycommfont}
\usepackage{lipsum}
\usepackage{epsfig} 
\usepackage{fancyhdr} 
\usepackage{pdflscape}
\usepackage{wasysym}
\usepackage{graphicx}
\usepackage{dashrule}
\usepackage{pgffor}
\usepackage{theorem}
\usepackage{listings} 
\usepackage{amssymb} 
\usepackage{pdfpages}
\usepackage{paralist}
\usepackage{changepage}
\usepackage{listings}
\usepackage{xspace}
\usepackage{wrapfig}
\usepackage{enumerate}
\usepackage{paralist}
\usepackage{url, polynom}
\usepackage[normalem]{ulem}
\usepackage{comment}
\usepackage{mwe}
\usepackage{multirow}

\usepackage[export]{adjustbox}
\usepackage{float}
\usepackage{enumitem}
\usepackage{array}
\usepackage{pdflscape}
\usepackage{graphicx}
\usepackage{boxedminipage}
\usepackage{enumitem}
\usepackage{afterpage}
\usepackage[T1]{fontenc}
\usepackage[utf8]{inputenc}
\usepackage[subtle]{savetrees}

\setlist{nolistsep,leftmargin=*}
\newlength\myindent
\newcommand{\java}{\textnormal{Java}\xspace}
\newcommand{\jml}{JML\xspace}
\newcommand{\aem}{{\AE}minium\xspace}
\usepackage{xcolor}
\definecolor{commentgreen}{RGB}{2,112,10}
\newcommand{\quotes}[1]{``#1''}
\newcommand{\jomp}{Java Grande benchmark\xspace}
\newcommand{\plaid}{Plaid\xspace}
\newcommand{\crystalsaf}{Crystal\xspace}
\newcommand{\plural}{Plural\xspace}
\newcommand{\javasyp}{JavaSyp\xspace}
\newcommand{\pulse}{Pulse\xspace}

\newcommand{\gap}{\texttt{Sip4J}\xspace}

\newcommand{\javasl}{\textnormal{Java Specification Language}\xspace}
\newcommand{\jvm}{\textnormal{Java Virtual Machine}\xspace}

\newcommand{\var}{\texttt{<grv>}\xspace}
\newcommand{\varnum}[1]{\texttt{<grv#1>}\xspace}
\newcommand{\lvar}{\texttt{<lv>}\xspace}

\newcommand{\localrefvar}{\texttt{<lrv>}\xspace}
\newcommand{\localrefvarnum}[1]{\texttt{<lrv#1>}\xspace}

\newcommand{\DbC}{\textnormal{Design by Contract principle}\xspace}

\newcommand{\foo}{\texttt{this\_m}\xspace}

\newcommand{\context}{\texttt{context}\xspace}
\newcommand{\variable}{\texttt{variable}\xspace}

\newcommand{\contextn}{\texttt{Context-N}\xspace}
\newcommand{\contextr}{\texttt{Context-R}\xspace}
\newcommand{\contextrw}{\texttt{Context-RW}\xspace}
\makeatletter
\newcommand{\xdashrightarrow}[2][]{\ext@arrow 0359\rightarrowfill@@{#1}{#2}}
\newcommand{\xdashleftarrow}[2][]{\ext@arrow 3095\leftarrowfill@@{#1}{#2}}
\newcommand{\xdashleftrightarrow}[2][]{\ext@arrow 3359\leftrightarrowfill@@{#1}{#2}}
\def\rightarrowfill@@{\arrowfill@@\relax\relbar\rightarrow}
\def\leftarrowfill@@{\arrowfill@@\leftarrow\relbar\relax}
\def\leftrightarrowfill@@{\arrowfill@@\leftarrow\relbar\rightarrow}
\def\arrowfill@@#1#2#3#4{%
  $\m@th\thickmuskip0mu\medmuskip\thickmuskip\thinmuskip\thickmuskip
   \relax#4#1
   \xleaders\hbox{$#4#2$}\hfill
   #3$%
}

\newcommand{\code}{\normalsize\texttt}
\newcommand{\equfont}{\footnotesize\texttt}

\newcommand{\ignore}[1]{}

\renewcommand{\arraystretch}{0.8}

\newcommand{\mbeq}{\overset{!}{=}} 

\newdimen\proofrulebreadth \proofrulebreadth=.05em
\newdimen\proofdotseparation \proofdotseparation=1.25ex
\newdimen\proofrulebaseline \proofrulebaseline=2ex
\newcount\proofdotnumber \proofdotnumber=3
\let\then\relax
\def\hfi{\hskip0pt plus.0001fil}
\mathchardef\squigto="3A3B
\newif\ifinsideprooftree\insideprooftreefalse
\newif\ifonleftofproofrule\onleftofproofrulefalse
\newif\ifproofdots\proofdotsfalse
\newif\ifdoubleproof\doubleprooffalse
\let\wereinproofbit\relax
\newdimen\shortenproofleft
\newdimen\shortenproofright
\newdimen\proofbelowshift
\newbox\proofabove
\newbox\proofbelow
\newbox\proofrulename
\def\shiftproofbelow{\let\next\relax\afterassignment\setshiftproofbelow\dimen0 }
\def\shiftproofbelowneg{\def\next{\multiply\dimen0 by-1 }%
\afterassignment\setshiftproofbelow\dimen0 }
\def\setshiftproofbelow{\next\proofbelowshift=\dimen0 }
\def\setproofrulebreadth{\proofrulebreadth}

\def\prooftree{
\ifnum  \lastpenalty=1
\then   \unpenalty
\else   \onleftofproofrulefalse
\fi
\ifonleftofproofrule
\else   \ifinsideprooftree
        \then   \hskip.5em plus1fil
        \fi
\fi
\bgroup
\setbox\proofbelow=\hbox{}\setbox\proofrulename=\hbox{}%
\let\justifies\proofover\let\leadsto\proofoverdots\let\Justifies\proofoverdbl
\let\using\proofusing\let\[\prooftree
\ifinsideprooftree\let\]\endprooftree\fi
\proofdotsfalse\doubleprooffalse
\let\thickness\setproofrulebreadth
\let\shiftright\shiftproofbelow \let\shift\shiftproofbelow
\let\shiftleft\shiftproofbelowneg
\let\ifwasinsideprooftree\ifinsideprooftree
\insideprooftreetrue
\setbox\proofabove=\hbox\bgroup$\displaystyle 
\let\wereinproofbit\prooftree
\shortenproofleft=0pt \shortenproofright=0pt \proofbelowshift=0pt
\onleftofproofruletrue\penalty1
}

\def\eproofbit{
\ifx    \wereinproofbit\prooftree
\then   \ifcase \lastpenalty
        \then   \shortenproofright=0pt  
        \or     \unpenalty\hfil         
        \or     \unpenalty\unskip       
        \else   \shortenproofright=0pt  
        \fi
\fi
\global\dimen0=\shortenproofleft
\global\dimen1=\shortenproofright
\global\dimen2=\proofrulebreadth
\global\dimen3=\proofbelowshift
\global\dimen4=\proofdotseparation
\global\count255=\proofdotnumber
$\egroup  
\shortenproofleft=\dimen0
\shortenproofright=\dimen1
\proofrulebreadth=\dimen2
\proofbelowshift=\dimen3
\proofdotseparation=\dimen4
\proofdotnumber=\count255
}

\def\proofover{
\eproofbit 
\setbox\proofbelow=\hbox\bgroup 
\let\wereinproofbit\proofover
$\displaystyle
}%
\def\proofoverdbl{
\eproofbit 
\doubleprooftrue
\setbox\proofbelow=\hbox\bgroup 
\let\wereinproofbit\proofoverdbl
$\displaystyle
}%
\def\proofoverdots{
\eproofbit 
\proofdotstrue
\setbox\proofbelow=\hbox\bgroup 
\let\wereinproofbit\proofoverdots
$\displaystyle
}%
\def\proofusing{
\eproofbit 
\setbox\proofrulename=\hbox\bgroup 
\let\wereinproofbit\proofusing
\kern0.3em$
}

\def\endprooftree{
\eproofbit 
  \dimen5 =0pt
\dimen0=\wd\proofabove \advance\dimen0-\shortenproofleft
\advance\dimen0-\shortenproofright
\dimen1=.5\dimen0 \advance\dimen1-.5\wd\proofbelow
\dimen4=\dimen1
\advance\dimen1\proofbelowshift \advance\dimen4-\proofbelowshift
\ifdim  \dimen1<0pt
\then   \advance\shortenproofleft\dimen1
        \advance\dimen0-\dimen1
        \dimen1=0pt
        \ifdim  \shortenproofleft<0pt
        \then   \setbox\proofabove=\hbox{%
                        \kern-\shortenproofleft\unhbox\proofabove}%
                \shortenproofleft=0pt
        \fi
\fi
\ifdim  \dimen4<0pt
\then   \advance\shortenproofright\dimen4
        \advance\dimen0-\dimen4
        \dimen4=0pt
\fi
\ifdim  \shortenproofright<\wd\proofrulename
\then   \shortenproofright=\wd\proofrulename
\fi
\dimen2=\shortenproofleft \advance\dimen2 by\dimen1
\dimen3=\shortenproofright\advance\dimen3 by\dimen4
\ifproofdots
\then
        \dimen6=\shortenproofleft \advance\dimen6 .5\dimen0
        \setbox1=\vbox to\proofdotseparation{\vss\hbox{$\cdot$}\vss}%
        \setbox0=\hbox{%
                \advance\dimen6-.5\wd1
                \kern\dimen6
                $\vcenter to\proofdotnumber\proofdotseparation
                        {\leaders\box1\vfill}$%
                \unhbox\proofrulename}%
\else   \dimen6=\fontdimen22\the\textfont2 
        \dimen7=\dimen6
        \advance\dimen6by.5\proofrulebreadth
        \advance\dimen7by-.5\proofrulebreadth
        \setbox0=\hbox{%
                \kern\shortenproofleft
                \ifdoubleproof
                \then   \hbox to\dimen0{%
                        $\mathsurround0pt\mathord=\mkern-6mu%
                        \cleaders\hbox{$\mkern-2mu=\mkern-2mu$}\hfill
                        \mkern-6mu\mathord=$}%
                \else   \vrule height\dimen6 depth-\dimen7 width\dimen0
                \fi
                \unhbox\proofrulename}%
        \ht0=\dimen6 \dp0=-\dimen7
\fi
\let\doll\relax
\ifwasinsideprooftree
\then   \let\VBOX\vbox
\else   \ifmmode\else$\let\doll=$\fi
        \let\VBOX\vcenter
\fi
\VBOX   {\baselineskip\proofrulebaseline \lineskip.2ex
        \expandafter\lineskiplimit\ifproofdots0ex\else-0.6ex\fi
        \hbox   spread\dimen5   {\hfi\unhbox\proofabove\hfi}%
        \hbox{\box0}%
        \hbox   {\kern\dimen2 \box\proofbelow}}\doll%
\global\dimen2=\dimen2
\global\dimen3=\dimen3
\egroup 
\ifonleftofproofrule
\then   \shortenproofleft=\dimen2
\fi
\shortenproofright=\dimen3
\onleftofproofrulefalse
\ifinsideprooftree
\then   \hskip.5em plus 1fil \penalty2
\fi
}

\patchcmd{\thebibliography}
  {\settowidth}
  {\setlength{
  \itemsep}{
    0pt plus 0.5pt}
  \settowidth}
  {}{}
\apptocmd{\thebibliography}
  {\footnotesize}
  {}{}

\begin{document}
\begin{frontmatter}
\cortext[cor1]{Corresponding author}
\title{Automatic Inference of Symbolic Permissions for Sequential Java Programs}
\author[rvt]{Ayesha Sadiq\corref{cor1}}
\ead{ayesha.sadiq@monash.edu}
\author[rvt]{Yuan-Fang Li}
\ead{yuanfang.li@monash.edu}
\author[rvt]{Li Li}
\ead{li.li@monash.edu}
\author[rvt]{Chris Ling}
\ead{chris.ling@monash.edu}
\author[rvt1]{Ijaz Ahmed}
\ead{ijaz.ahmed@cs.uol.edu.pk}
\address[rvt]{Faculty of Information Technology, Monash University Victoria, Australia}
\address[rvt1]{Department of Computer and Information Sciences, University of Lahore, Pakistan}
\begin{abstract}

In mainstream programming languages such as \java, a common way to enable concurrency is to manually introduce explicit concurrency constructs such as multi-threading. In multi-threaded programs, managing synchronization between threads is a complicated and challenging task for the programmers due to thread interleaving and heap interference that leads to problems such as deadlocks, data races. With these considerations in mind, access permission-based dependencies have been investigated as an alternative approach to verify the correctness of multi-threaded programs and to exploit the implicit concurrency present in sequential programs without using explicit concurrency constraints. 
However, significant annotation overhead can arise from manually adding permission-based specifications in a source program, diminishing the effectiveness of existing permission-based approaches. 

In this paper, we present a framework, \gap, to automatically extract access permission-based implicit dependencies from sequential \java programs, by performing inter-procedural static analysis of the source code. Moreover, we integrate and extend an existing permission-based verification tool, \pulse, to automatically verify correctness of the inferred specifications and to reason about their concurrent behaviors. Our evaluation on some widely-used benchmarks gives strong evidence of the correctness of the inferred annotations and their effectiveness in enabling concurrency in sequential programs.

\end{abstract}
\begin{keyword}
Access permissions \sep permission inference \sep static analysis \sep
object oriented programs \sep \java
\end{keyword}
\end{frontmatter}
\linenumbers

\section{Introduction}

Enabling concurrency for imperative and object-oriented languages has become one of the grand challenges for the IT industry today\footnote{UK Computing Research Committee, Grand Challenges in Computing Research. \url{http://www.ukcrc.org.uk/grand-challenge/.}}. This is because of the nature of the imperative and object-oriented programming paradigms where the compiler follows the execution order in which the program is written, i.e., sequential.
In such languages, programmers manually introduce concurrency by using explicit concurrency constructs, e.g., multi-threading-related classes such as \emph{Thread}, \emph{Runnable} in Java.
Unfortunately, traditional multi-threading models frequently result in deadlocks or unwarranted race conditions that are hard to debug.
Fractional permissions, a way to ensure the non-interference of program states in parallel programs
is hence introduced to alleviate these issues \cite{boyland2003checking}. Fractional permissions encode the read and write effects of a reference on a referenced object using concrete values 1 and 0 where the value 0 represents the absence of permission, whereas 1 represents \code{unique} (exclusive read and write) permission on the referenced object and any value greater than zero models the read-only access for a shared object.

Symbolic permission \cite{bierhoff2007modular,Beckman:2008}, simply called \emph{access permission}, is an extension of fractional permissions. It is a novel abstraction that combines the read, write and aliasing information of a referenced object. Instead of using fractional values to represent the read and write access among multiple references, symbolic permissions represent and track permission flow through the system using high-level abstractions (permission types) such as \textsf{immutable} or \textsf{unique}, etc. 

A group of CMU researchers led by Jonathan Aldrich manually wrote permission-based typestate contracts on a number of \java APIs to model and reason about the correctness of usage protocols in typestate-based sequential and concurrent programs \cite{bierhoff2007modular,Bierhoff-plural-tool,Beckman:2008,beckman2009modular,Bierhoff09polymorphicfractional} etc., and further parallelise the execution of these programs in a permission-based typestate-oriented programming paradigm \plaid ~\cite{aldrich2011plaid}. Further, having the access permission support in the \plaid infrastructure, the group worked in a joint research project and designed a by-default concurrent programming language and a runtime system, \aem~\cite{Aeminium2014}, to parallelise execution of sequential programs based on access permissions.
Furthermore, inference of fractional and quantified permissions has been investigated by Peter M{\"{u}}ller and his colleagues \citep{Ferrara2012,Dohrau2018PermissionPrograms} to verify class-based concurrent programs based on the abstract interpretations \cite{Cousot1977}.
Similarly, permission-based specifications have been used in many formal approaches to address issues related to safe concurrency, security and verification of functional and domain specific properties \cite{Chalice2009,Le2012,Boyland2014ConstraintPermissions,Wickerson2010ExplicitReasoning,jacobs2011verifast,SiminiceanuAC12,CatanoASA14,Juhasz2014Viper,
Huisman2015AReasoning,Muller2017Viper:Reasoning,Jacobs:2018}.



Unfortunately, in order to benefit from access permissions, programmers have to manually add appropriate permission-based specifications (e.g., annotations) as dependency information in the program. Not only do programmers need to spend time getting familiarised with a completely new specification language and runtime system, they also need to manually write specifications in the source code which is laborious and error-prone. Moreover, given the intricacies in creating these constructs, it is very likely for a programmer to omit important dependencies or create wrong specifications that may again lead to problems such as race-conditions and (or) deadlocks. These issues have hindered the wider adoption of access permission-based approaches.

To this end, in our work, we aim to resolve the aforementioned issues by introducing to the community a novel approach that infers implicit dependencies from the source program in the form of access permissions 
The goal is to free programmers from
the annotation overhead for manually adding permission-based dependencies in
the program, thereby solving the common problem faced by the existing access
permission-based approaches.



We are interested in inferring permission-based dependencies for \java programs. This paper presents a comprehensive framework called \gap to infer access permission contracts from sequential \java programs using permission inference technique. To infer permission-based dependencies from the source code, our technique is based on Abstract Syntax Tree and follows a set of pre-defined syntactic rules and graph abstractions. A tool based on the technique is implemented as an Eclipse Plugin, integrated with an existing permission-based model-checking tool \pulse~\cite{SiminiceanuAC12}. We then empirically evaluate the proposed technique by verifying the correctness of the inferred specifications and hence demonstrating the effectiveness of the tool. We also measure the execution speed of our inference technique on a number of widely-used benchmark programs and realistic \java applications.

It is worth mentioning that the technique proposed in this work, although focused on \java language only, should also be applicable to other object-oriented programming languages. Indeed, we believe that the inferred specifications and our framework can be used
by existing permission-based verification approaches to verify the correctness of a program without annotation overhead. It is also able to discover some of the syntactical errors in a program such as null pointer references at compile time. Ultimately, it can be used to reason about the concurrent behaviour of a sequential program without imposing extra work on the programmers.

To summarise, in this work, we make the following contributions:
\begin{itemize}
    \item We design and implement a prototype tool called \tool{} for automatically inferring permission-based specifications for sequential Java programs. Additionally, the core functionalities of \tool{} are further integrated into Eclipse as a plugin.
    
    \item We integrate and extend Pulse to automatically verify the validity of permission specifications generated by \tool{} and to reason about their concurrent behavior.
    
    \item We evaluate \tool{} on both benchmark and real-world sequential Java programs. Our experimental results show that \tool{} is indeed capable of inferring access permissions for Java programs. Furthermore, with additional experiments, we empirically show that the inferred permissions are not only useful for parallelising Java programs but also helpful for pinpointing program errors such as null pointer references and to perform reachability (code) analysis of the program at method level.
\end{itemize}

To help readers replicate our results, we have provided the \gap tool, the benchmark programs and their analysis reports by the \gap framework on GitHub: 

\begin{center}
    \url{https://github.com/Sip4J/Sip4J}
\end{center}

The rest of the paper is structured as follows. Section \ref{prelim} provides an introduction to access permissions, permission contracts and a motivating example. Section \ref{framework} describes the \gap framework to generate and verify permission-based specifications. Section \ref{sec:method} discusses the \gap permission inference approach in detail and Section \ref{perm-checker} discusses \pulse as a permission checker and its integration within the \gap framework. Section \ref{evaluation} demonstrates the correctness and the effectiveness of the proposed technique on benchmark applications. 
In Section \ref{implications} and \ref{threats}, we elaborates on the implications and limitations of the \tool{} framework and its analysis in general. Section \ref{relatedwork} discusses the related work. Finally, we conclude the paper in Section \ref{conclusion} and propose some future directions.

\section{Motivation and Preliminaries}
\label{prelim}

This section briefly explains and elaborates the challenges posed by parallelising imperative programming models and how access permission-based specifications can be used to handle those challenges.

\subsection{Parallelising sequential imperative programs: A motivation example}
Multi-core architectures are ubiquitous and have become the norm due to the parallel execution power made available by the multi-core processors in them.
Herb~\cite{Sutter} famously stated ``\emph{The free lunch is over}'' which means current applications can no longer benefit from the free ride (rapid performance improvements) unless programmers redesign current applications or otherwise exploit the potential concurrency present in the system.  However, software industry has not yet been able to fully exploit the performance boost of multi-core systems, principally due to inherit limitations of mainstream (imperative and object-oriented) programming languages such as Java, C++, etc. 
Indeed, most of the existing applications are still being written in sequential programming paradigms, without using multi-threading, which cannot benefit from the characteristics of multi-core machines.

In order to benefit from modern multi-core systems, there is a need to transform traditional sequential programs to parallel programs, so as to improve the execution time of these programs and to free programmers from the low level ordering and reasoning overhead about thread synchronization.

Unfortunately, in imperative programming languages such as \java, because of the implicit dependencies between the code and shared states, methods dependant on the same mutable object do not explicitly show their side effects (the read and write behaviors) to each other. It is hence non-trivial for programmers to manually parallelise sequential programs without the fear of data races. 
Let us take Listing~\ref{lst:ua-specprog} as an example.

\begin{lstlisting}[caption={A sample Java program.},label=lst:ua-specprog,firstnumber=1]
class ArrayCollection{
 public Integer[] array1;
 ArrayCollection(){
 array1 = new Integer[10];
 for(int i = 0; i < array1.length; i++)
  array1[i] = (int)(Math.random() * 10);
 }
 public void printColl(Integer[] coll) {
  for(int i = 0; i < coll.length; i++){
   System.out.println(" "+coll[i]);}
 }
 public  void incrColl(Integer [] coll) {
  for (int i = 0; i < this.array1.length; i++){
    this.array1[i] = this.array1[i] + i; }
  for (int j = 0; j < coll.length; j++){
    coll[j] = coll[j] + j; }
 }
 public boolean isSorted(Integer[] coll) {
 boolean flag = false;
 int j = 0;
 for(int i = 0;i<coll.length && j<coll.length;i++){
  if(coll[i] > coll[j])
   flag = true;
  else
   flag = false;
  }
  return flag;
 }
 public Integer findMax(Integer[] coll) {
  int max;
  max = coll[0];
  for (int i = 1; i < coll.length; i++){
   if(coll[i] > max)
     max = coll[i];}
  return max;
 }
 public void computeStat(Integer[] coll){
  printColl(coll);
  System.out.println("Is sorted = "+isSorted(coll));
  System.out.println("Max number = "+findMax(coll));
 }
 public void tidyupColls(Integer[] coll){
    this.array1 = null;
    coll = null;}
}
class ObjectClass{
 public Integer[] array2;
 public  Client x,y,z,w;
 ObjectClass(){
  array2 = new Integer[10];
  for(int i = 0; i < array2.length; i++){
   array2[i] = (int)(Math.random() * 10);
  }
  x = new Client(); y = new Client(); z = new Client(); w = new Client();
 }
 public void manipulateObjects(Client p1, Client p2){
   x = p1;
   Client t = x;
   y = t;
   x.data = 10;
   System.out.println("z.data = "+p2.data;}
}
class Client{	
 Integer data = 100;
 public static void main(String[] a) {
  ArrayCollection obj1 = new ArrayCollection();
  ObjectClass obj2 = new ObjectClass();
  obj1.incrColl(obj2.array2);
  obj1.computeStat(obj1.array1);
  obj1.computeStat(obj2.array2);
  obj1.tidyupColls(obj2.array2);
  obj2.manipulateObjects(obj2.w, obj2.z);
 }
}
\end{lstlisting}

Listing \ref{lst:ua-specprog} illustrates a sequential \java program with three user-defined classes \code{ArrayCollection}, \code{ObjectClass} and  \code{Client}. These three classes contain eight methods and access two shared collection objects (i.e., \code{array1} and \code{array2}).
As an example, the \code{ObjectClass} composes the \code{Client} class and manipulates its member \code{data} using method \code{manipulateObjects()}. 
To benefit from the modern multi-core facilities,  certain methods of this sample program can be parallelly executed so as to improve the overall performance.
Indeed, if a given two methods do not write the same object at the same time, these two methods could be parallelised.
For example, since methods \code{computeState(array1)}, \code{computeState(array2)} and \code{manipulateObjects()} do not manipulate the same shared objects, they could be potentially parallelised. Oppositely, methods \code{incrColl(array2)} and \code{tidyUpColl(array2)} can not be parallelised as they may write the same object at the execution time.

Unfortunately, manually identifying and tracking the object accesses and the order in which these accesses are made is a laborious (in-terms of time and effort) and error-prone task for programmers. It is very likely for a programmer to omit important dependencies or identify the wrong dependencies. 
The situation becomes worse in case of unrestricted aliasing in the program, the hallmark feature of imperative programming models \cite{BierhoffBA09}.
Indeed, analysis of method \code{manipulateObjects()} shows that it accesses four objects (\code{x, y, z, w}) as its data members. Explicitly, it mutates only one object i.e. \code{x} by writing on its \code{data} field but actually it mutates object \code{y} and \code{w} as well due to aliasing of the objects withing the method.
These alias variables will create side effects for other methods accessing the same objects when executed in parallel. 
Moreover, as the program size (complexity) increases, it becomes non-trivial for a programmer to identify the implicit dependencies between different program parts by just looking at the source code. 

To exploit the potential concurrency present in a \java program, every method should either avoid these side effects (change in some value outside method scope) or should explicitly mention them. This information can be used to compute the data dependencies at different level of granularity within the code and parallelise execution of the program to the extent permitted by these dependencies.
Therefore, we need a mechanism that can express the mutability and aliasing between shared objects that can mitigate the undesirable effects to other methods in an expressive way, while allowing a safe execution order within methods. The objective is to free programmers from the tedious and low-level analysis overhead for identifying and tracking the implicit dependencies present at the code level. 

The study of literature shows that access permissions, a novel abstraction that combines (models) the effects (read and write) and aliasing information of a referenced object, provides a flexible control mechanism to track all the references of a particular object and update state changes to all such references~\citep{bierhoff2007modular}.
Access permission is hence suitable for characterizing the way (read and write) a shared resource is accessed by multiple references and can express the implicit dependencies present in the system while making them explicit, thereby the side effects.
Furthermore, permission-based specifications pose their own ordering constraints and therefore can be used to perform method's operations in a non-interfering manner and parallelise code without using the low-level concurrency and ordering constraints \cite{aldrich2012plaid,Aeminium2014}.

\subsection{Access Permissions}\label{sec:accessperm}

Before elaborating the efficacy and expressiveness of permission-based specifications to model aliasing and parallelising sequential imperative programs, we provide an overview of access permissions semantics, permission co-existence, access permission splitting (joining) rules and permission-based contracts.

Access permissions are used to describe whether or not an object is being aliased, whether a given reference can modify the referenced object, and whether other references (aliases) that point to the same object, if any, are allowed to modify the object \citep{bierhoff2007modular}.

Let \textsf{x} and \textsf{y} be the current and other reference respectively and let \textsf{o} represent a referenced object. 
There are five (symbolic) permission types that can be assigned to a reference \textsf{x} for the referenced object \textsf{o} in the presence of the alias \textsf{y}.

\begin{description}
\item[\textbf{unique(x)}]: This permission provides to reference \textsf{x} an exclusive read and modify access on the referenced object \textsf{o} at any given time. No other reference (e.g.\ \textsf{y}) to the same object can co-exist while \textsf{x} has unique permission on \textsf{o}.\\

\begin{minipage}{.4\textwidth}
\centering\includegraphics[scale=0.6]{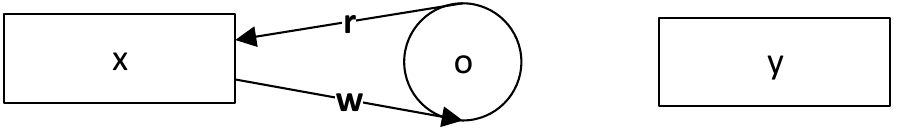}
\end{minipage}
\\
\item[\textbf{full(x)}]: This permission grants reference \textsf{x} with read and write access to the referenced object \textsf{o}, and at the same time \textsf{o} may also be read, but not written, by other reference \textsf{y}.\\

\begin{minipage}{.4\textwidth}
\centering\includegraphics[scale=0.6]{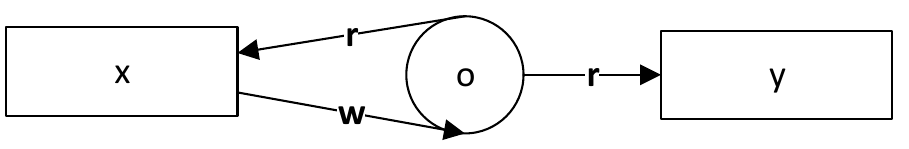}
\end{minipage}
\\
\item[\textbf{share(x)}]: This permission is the same as \textsf{full(x)}, except that now other references  \textsf{y} can also write on the referenced object \textsf{o}.\\

\begin{minipage}{.4\textwidth}
\centering\includegraphics[scale=0.6]{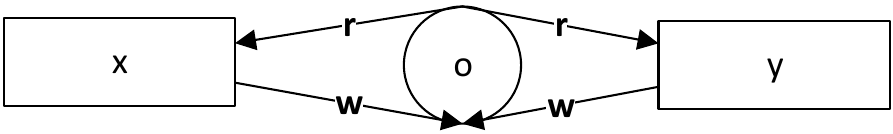}
\end{minipage}
\\
\item[\textbf{pure(x)}]: This permission gives a reference \textsf{x} read-only access on a referenced object \textsf{o}. Moreover, other reference \textsf{y} may have read and write access on the same object.\\

\begin{minipage}{.4\textwidth}
\centering\includegraphics[scale=0.6]{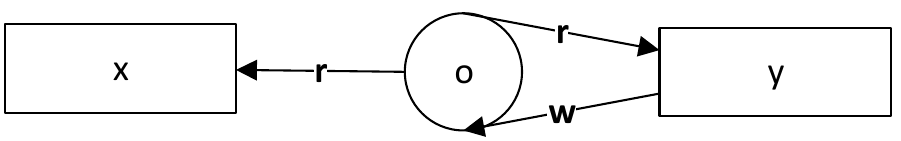}
\end{minipage}
\\
\item[\textbf{immutable(x)}]: This permission grants a non-modifying access on the referenced object \textsf{o} to both the current reference \textsf{x} and any other reference \textsf{y}.\\

\begin{minipage}{.4\textwidth}
\centering\includegraphics[scale=0.6]{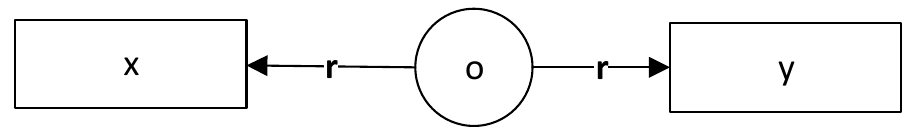}
\end{minipage}
\end{description}

Table \ref{table:Table1} below summarises how access permissions on a referenced object \textsf{o} can co-exist between the \emph{current reference} \textsf{(x)} and the \textit{other reference} \textsf{(y)} \cite{Boyland2013FractionalPermissions}. 
\begin{table}[H]
\centering
\caption{Co-existing access permissions on the same object~\cite{boyland2003checking}}
\label{table:Table1}
\resizebox{\linewidth}{!}{
\begin{tabular}{lll}
\hline\noalign{\smallskip}
\textbf{This reference \textsf{(x})} &\textbf{Access rights of \textsf{(x)}}& \textbf{Other references (y)} \\
\noalign{\smallskip}\hline\noalign{\smallskip}
\textsf{unique}&read/write&none\\
\textsf{full}&read/write&\textsf{pure}\\
\textsf{share}&read/write&\textsf{share}, \textsf{pure}\\
\textsf{pure}&read&\textsf{full}, \textsf{pure}, \textsf{immutable}\\
\textsf{immutable}&read&\textsf{immutable}, \textsf{pure}\\
\noalign{\smallskip}\hline
\end{tabular}
}
\end{table}
The table shows that the \textsf{unique} access permission is very restrictive, as it consumes all the read and write permissions on a particular object and does not share any access with any other reference on the same object. In contrast, \textsf{pure} is the least restrictive as it only consumes read permission and at the same time shares read permission and gives other references the exclusive write access on the same object.

To infer five types of symbolic permissions from the source code of a \java program, our permission inference approach models the object's accesses and their aliasing information at the method level, in the form of permission-based graph models as shown above. In the generated models, \code{x} corresponds to the current method and \code{y} corresponds to other methods ("the rest of the world except the current method") accessing the same object \code{o}. The details of the permission inference and the graph construction mechanism are further explained in Section \ref{framework}.

\subsubsection{Access permission splitting and joining rules}\label{permRules}
Access permissions are considered as resources in Linear logic \citep{girard1987linear} that cannot be duplicated (discarded). Once a method consumes its permissions they are no longer available to other methods until this method returns the same permissions again. 

Access permission can be split into one or more relaxed permissions i.e., fractions of original permission using fractional values in the range $(0,1)$) and then merged back into more restrictive or original permission. This phenomenon is known as fractional permission analysis where fractions keep tracks of the way the permissions were split and joined back. This information can be used to verify system properties, based on specific criteria and parallelise execution of the program by tracking the permission flow through the system.

Table \ref{table:Table2} shows access permissions splitting and joining rules to track permission flow through the program.

\begin{table}[!t]
\begin{center}
\caption{Access permissions splitting and joining rules \citep{bierhoff2007modular}.}
\label{table:Table2}
\resizebox{\linewidth}{!}{
\begin{tabular}{l}
\toprule
\textbf{Permission Splitting and joining Rules} \\
\hline
1.  unique(x;o;k) $\Leftrightarrow$ full($x_{1}$;o;$k_{1}$) $\bigotimes$ pure($x_{2}$;o;$k_{2}$)\\
2.  unique(x;o;k) $\Leftrightarrow$	immutable($x_{1}$;o;$k_{1}$)$\bigotimes$ immutable($x_{2}$;o;$k_{2}$)\\
3.  full(x;o;k) $\Leftrightarrow$ share($x_{1}$;o;$k_{1}$) $\bigotimes$ pure($x_{2}$;o;$k_{2}$)\\
4.  share(x;o;k) $\Leftrightarrow$ full($x_{1}$;o;$k_{1}$) $\bigotimes$ pure($x_{2}$;o;$k_{2}$)\\
5.  immutable(x;o;k) $\Leftrightarrow$ pure($x_{1}$;o;$k_{1}$)$\bigotimes$ immutable($x_{2}$;o;$k_{2}$)\\
6.  unique(x;o;k) $\Leftrightarrow$	share($x_{1}$;o;$k_{1}$)$\bigotimes$ share($x_{2}$;o;$k_{2}$)\\
7.  immutable(x;o;k) $\Leftrightarrow$ immutable($x_{1}$;o;$k_{1}$) $\bigotimes$ immutable($x_{2}$;o;$k_{2}$)\\
8.  share(x;o;k) $\Leftrightarrow$ share($x_{1}$;o;$k_{1}$) $\bigotimes$ pure($x_{2}$;o;$k_{2}$)\\
9.  share(x;o;k) $\Leftrightarrow$ share($x_{1}$;o;$k_{1}$) $\bigotimes$ share($x_{2}$;o;$k_{2}$)\\
10. full(x;o;k) $\Leftrightarrow$ full($x_{1}$;o;$k_{1}$) $\bigotimes$ pure($x_{2}$;o;$k_{2}$)\\
\bottomrule
\end{tabular}}
\end{center}
\end{table}

In Table 2, $x$ represents current reference, \code{o} represents the shared object and $k$ represents the fraction of permission assigned to a particular reference 
The operator multiplicative conjunction (A $\bigotimes$ B) denotes simultaneous occurrence of permissions by multiple references, say $x_{1}$, $x_{2}$, on the same referenced object \code{o}. The symbol $\Leftrightarrow$ represents the two way operation of splitting and joining permissions. 
For example, \textsf{unique} access permission (Rule 1.) having \code{k} fractions can be divided into $k_{1}$ fraction of \textsf{full} and $k_{2}$ fraction of \textsf{pure} permission on the same object and then joined back accordingly. Likewise, a \textsf{unique} access permission (Rule 6.) can be split into two \textsf{share} permissions but cannot be split into a \textsf{share} and \textsf{immutable} permission as the \textsf{immutable} permission cannot co-exist with the \textsf{share} permission on the same object. Linearity of resources forces the \textsf{unique} permission to be consumed, replaced by two \textsf{share} permissions, which can be further split according to splitting rules and then joined back.

\subsubsection{Access Permission Contracts in Plural and Pulse}\label{plural-background}
In our work, we are interested in inferring permission-based specifications for \java programs. To the best of our knowledge, there are only two research tools that are directly related to Java-based access permissions, namely \plural\footnote{Pluralism: Modular Object Protocol Checking for Java, \url{https://code.google.com/archive/p/pluralism/}}\cite{Bierhoff-plural-tool} and \pulse\footnote{PULSE: A Model-Checking Tool to Verify Typestates and Access Permissions Specifications of Java Programs, \url{http://poporo.javerianacali.edu.co/aeminium/pulsepulse/pulse.php.}}\cite{SiminiceanuAC12}.

It is worth mentioning here that the permission inference approach, presented in this paper, integrates the \pulse tool to verify the correctness of the inferred specifications. It further extends the concurrency analysis in the \pulse tool to perform a comprehensive concurrency analysis of the input sequential program, based on five types of symbolic permissions. Therefore, it is non-trivial for readers of the permission inference approach to first understand the \plural specifications and \pulse analysis itself.

\pulse\cite{SiminiceanuAC12,CatanoASA14} is a permission-based model-checking tool implemented as \java Eclipse plug-in. It takes a \plural annotated program i.e., a \java program annotated with access permission contracts and typestate information, as input. The typestate information is added, as a part of method specifications, for the \pulse input program. However, inferring (verifying) the typestates information is not an immediate objective of this research. \pulse translates the \plural specifications into a semantically equivalent
abstract state-machine model. The model captures the dynamic behavior of a program as a sequence of method calls obeying the access permission semantics, and the typestate information associated with the referenced objects. \pulse employs the \textsf{evmdd-smc} symbolic model-checker \citep{RouxS10} to verify the machine model. The model checker ensures that the input specifications satisfy a set of core integrity properties , specified as Computation Tree Logic (CTL)\citep{CTL-logic} formulae, by performing the reachability graph analysis of the generated state space.

The permission checking approach employs \pulse analysis to verify the correctness of input specifications and to reason about their concurrent behavior in following ways.

\begin{description}[leftmargin=0cm]
\setlength{\itemsep}{4pt}

\item \textbf{Access permission Correctness.} \pulse enforces that access permissions do not violate their intended semantics, by generating he discrete state semantics of the input specifications and by encoding the generated semantics into a fully discrete machine model. It then verifies the correctness of the input specifications, by following the pre-defined CTL formulae through the model checking mechanism.

\item \textbf{Method Satisfiability Analysis.} \pulse identifies missing specifications by performing the method (un)satisfiability analysis of the input specifications based on the pre-defined CTL formulae. The method satisfiability (reachability) analysis of a method is based on the \code{requires} clause (pre-permission) in a method contract. The method satisfiability analysis checks whether the \code{pre-condition} of a method is met. A method is satisfied (reachable) if all its pre-conditions are met or if it obtains enough (pre-) permission to start its execution.
The presence of the unsatisfiable methods, due to the method's unsatisfiable pre-condition, indicates an error (misspelled or missing specifications) in the input specifications, or it can be due to program error such as the use of \code{null} references. The presence of an unsatisfiable method also indicates that no possible client can fulfill the method's contract i.e., the \code{requires} clause and this method is not called under any circumstances; thus the method remains unreachable.

\item \textbf{Concurrency Analysis.}
\pulse identifies whether two methods can be executed in parallel by identifying the immutable methods, i.e., methods that require read-only access on a shared object or the methods that should always run in isolation, by following the (pre-) permission contracts between two methods. It computes the number of methods that can be executed with at least one other method, at the class level, including the method concurrency analysis with itself.
\end{description}

Our permission inference framework borrows the \plural's syntax to generate the \pulse input program. Lets us briefly discuss the syntax and semantics of access permissions in the \plural specification language. A complete discussion of the \pulse correctness and its extended concurrency analysis, for the example program given in Listing \ref{lst:ua-specprog}, is explained in Section \ref{perm-checker}.

\plural (Permissions Let Us Reason about Aliases) \citep{bierhoff2007modular,Bierhoff-plural-tool} is a formal specification language and a tool, originally developed to ensure protocol compliance in typestate-based sequential programs such as \java APIs. The program verification in \plural is based on access permissions and typestate information where access permissions encodes the read, write and aliasing behaviour of a method on the referenced objects and typestates describe the set of valid object's states a method can be called on \cite{strom1986typestate}. \plural supports five types of symbolic permissions such as \textsf{unique}, \textsf{immutable}, \textsf{full}, \textsf{pure} and \textsf{share} as a part of method specifications. \plural follows the \DbC \citep{Meyer1988Object-OrientedConstruction} to specify access permission contracts, as pre- and post-conditions, at the method level. 



In a \plural program, the annotation \code{@Perm} is used to specify a permission contract where pre- and post-conditions are defined using \code{requires} and \code{ensures} clause respectively. A typestate in \plural is declared using \code{@State} clause.
Typestate `\code{alive}' is a default global (root) state an object can be in. The precondition (\code{requires} clause) in a method contract specifies the type of access permissions (\code{AP}), a method requires on a referenced object \code{(this)} and the typestate (\code{ts}), a referenced object should be in before the method starts its execution. 

The permission (\code{$ap$}) on a parameter is represented using notation \code{AP(\#i)} where \code{i} is an integer that maps the position of a parameter in a method declaration as $0 \to N-1$ where \code{N} is the number of parameters in a method signature. The post-condition (\code{ensures} clause) in a method contract specifies the permissions (\code{$AP\prime$}) that the method generates on the referenced object (the parameter) to return it back to the caller method and the typestate (\code{$s\prime$}) an object should hold when the method completes its execution. The symbol $\ast$ shows the multiplicity (one or more) of the referenced objects with permission annotations. The notation \code{ENDOFCLASS} in is used to distinguish multiple classes, in a source program, to perform the correctness analysis of the access permission contracts at class level by \pulse. 



\subsection{Access permission contracts: motivation example revisited}\label{conc-analysis}

We now revisit the motivating example we presented at the beginning of this section to demonstrate the efficacy and expressiveness of permission-based specifications in enabling implicit concurrency present in a sequential program.

Listing \ref{lst:sa-specprog} shows an annotated version of the sample \java program given in Listing \ref{lst:ua-specprog}, with permission contracts at the field level in \plural style using a single typestate '\code{alive}'. 
As discussed previously in Section \ref{plural-background}, the approach
adds the typestate (\code{alive}) information to perform the evaluation of the inferred specifications
by the \pulse tool, otherwise, inferring and verifying typestate is not an objective of this research.

For simplicity, for parameters, we refer the objects accessed against parameters using their identifier to show the aliasing information explicitly, instead of using the parameter notations as in \plural.

\begin{lstlisting}[label=lst:sa-specprog, mathescape, caption={Access permission contracts for the example program given in Listing \ref{lst:ua-specprog}},morecomment={[s][\color{violet}]{@}{( }}]
class ArrayCollection{
 public  Integer[] array1 = new Integer[10];
 @Perm(ensures="unique(array1) in alive")
 public ArrayCollection() {}

 @Perm(requires="pure(array1) in alive",
       ensures="pure(array1) in alive")
 public void printColl(Integer[] coll) {}

 @Perm(requires="share(array1)  in alive * share(array2) in alive",
       ensures="share(array1) in alive * share(array2) in alive")
 public void incrColl(Integer[] coll) {}

 @Perm(requires="pure(array1) in alive",
       ensures="pure(array1) in alive")
 public boolean isSorted(Integer[] coll) {}

 @Perm(requires="pure(array1) in alive",
       ensures="pure(array1) in alive")
 public Integer findMax(Integer[] coll) {}

 @Perm(requires="pure(array1) in alive",
       ensures="pure(array1) in alive")
 public void computeStat(Integer[] coll){}

 @Perm(requires="unique(array1)  in alive * unique(array2) in alive",
       ensures="none(array1) in alive * none(array2) in alive")
 public void tidyupColl(Integer[] coll){}
}
ENDOFCLASS
class ObjectClass{
 public Integer[] array2 = new Integer[10];
 public  Client x = new Client(), y = new Client();
 public  Client z = new Client(),w = new Client();
 @Perm(requires="full(x) * full(y) * full(w) * immutable(z) in alive",
 ensures="full(x) *  full(y) * full(w) * immutable(z) in alive")
 void manipulateObjects(Client p1, Client p2){}
}
ENDOFCLASS
class Client{	
 Integer data = 100;
 @Perm(requires="none(obj1) * none(obj2) in alive",
 ensures="unique(obj1) * unique(obj2) in alive ")
 public static void main(String[] a) {
  ArrayCollection obj1 = new ArrayCollection();
  ObjectClass obj2 = new ObjectClass();
  obj1.incrColl(obj2.array2);
  obj1.computeStat(obj1.array1);
  obj1.computeStat(obj2.array2);
  obj1.tidyupColls(obj2.array2);
  obj2.manipulateObjects(obj2.w,obj2.z);
}
}
ENDOFCLASS
\end{lstlisting}

Consider the permission contract for method \code{printColl()} in Listing \ref{lst:sa-specprog} at Line 6 \& 7, \quotes{\code{@Perm(requires="pure(array1) in alive", ensures="pure(array1)" in alive)}}. This contract, for method \code{printColl()}, is generated following the method call expression \code{obj1.computeStat(obj1.array1)} in Listing \ref{lst:ua-specprog} at Line 69. It states that the method needs \code{pure} permission, as pre-permission, on the object referenced by variable \code{array1}. The post-permission specifies that the method guarantees to return the consumed permissions, on the same object, to the caller of the method.

Having such specifications (permission-based dependencies) in an explicit way, 
one can not only compute the number of immutable (independent) methods i.e., the methods that do not change (access) a shared object, or the methods that should always run sequentially but can also define the way (order), these methods can be executed in parallel by considering methods side-effects. 

\begin{figure}[!t]
\includegraphics[width=\linewidth]{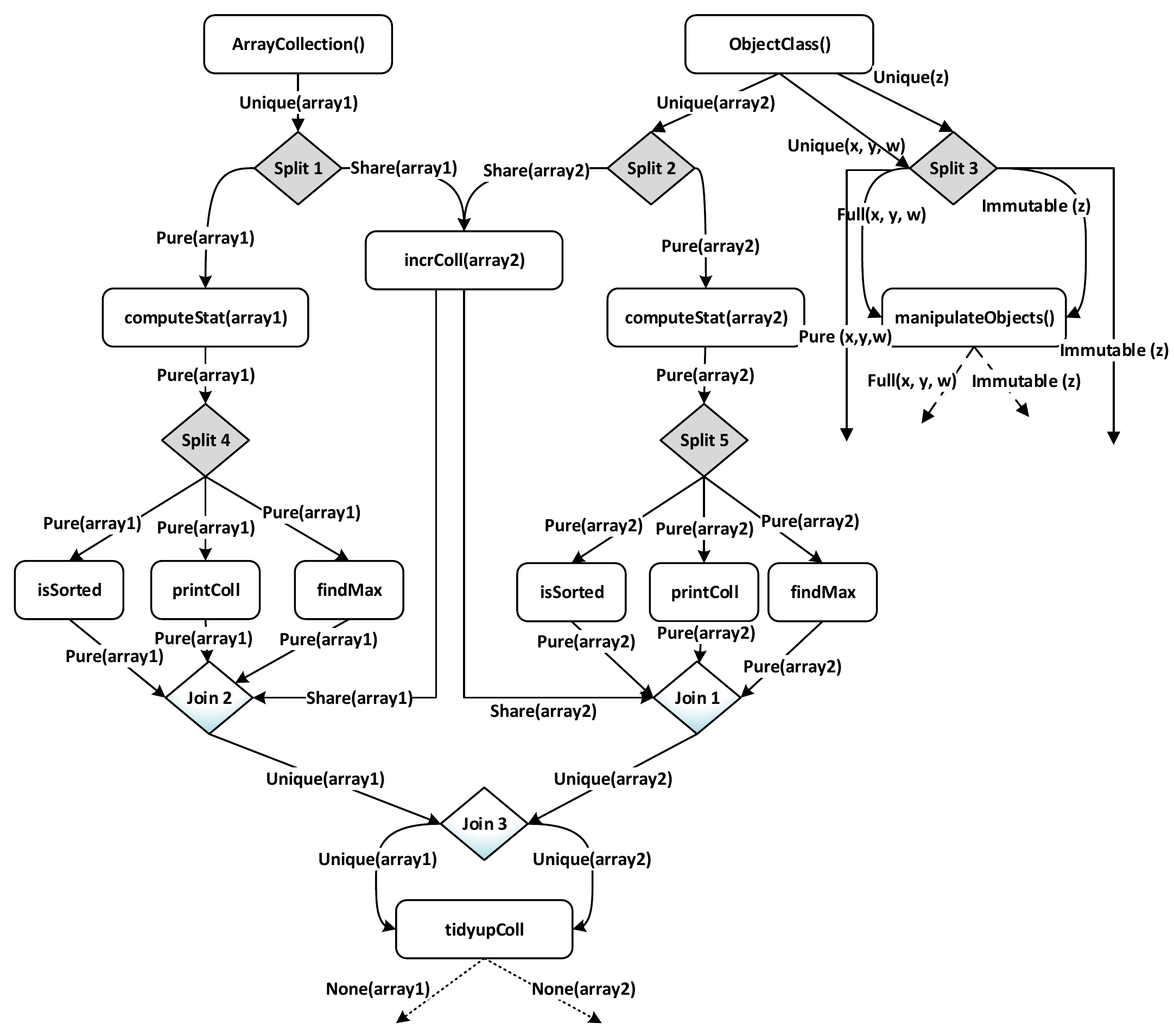}
\caption{Permission-based method (call) concurrency graph of the Java program in Listing \ref{lst:sa-specprog}.}
\label{fig:dependencygraph-example3}
\end{figure}

Figure \ref{fig:dependencygraph-example3} shows the method call concurrency graph of the annotated program following the access permission semantics (Section \ref{sec:accessperm}) and the access permission splitting and joining rules in Table \ref{table:Table2}. The analysis reveals that, in total, 11 out of 13 method calls across three classes can be executed in parallel.

The analysis shows that constructor methods cannot be parallelized with any method as no other method can use an object before it is being created with a \textsf{Unique} permission. However, constructors instantiating different objects can be executed at the same time. Further, it shows that methods that requires either read-only permissions (\textsf{pure} or \textsf{immutable}) such as \code{printColl()} and \code{findMax()} on the same (different) object can be executed in parallel. 

However, methods that require \textsf{full} (write) permission on a shared object e.g.\ \code{incrColl()} cannot be executed in parallel with other methods writing on the same object, due to the side effects they produce. However, these method can be parallelized with 
a) other methods that either require read (\textsf{pure}) or write (\textsf{full}) permission on different objects such as method \code{manipulateObjects()} and computeState() accessing different objects. Similarly, methods that requires \code{unique} permission on the referenced objects such as \code{tidyupColls()} should always run in isolation. 


However, manually adding permission-based specifications in a program pose significant annotation overhead for programmers and that is itself a tedious and error prone task to do by hand. Hence, the automatic inference of permission-based dependencies from the source code is desirable.

\section{Sip4j: The Permission Inference Framework}
\label{framework}

\begin{figure}[H]
\centering\includegraphics[width=\linewidth]{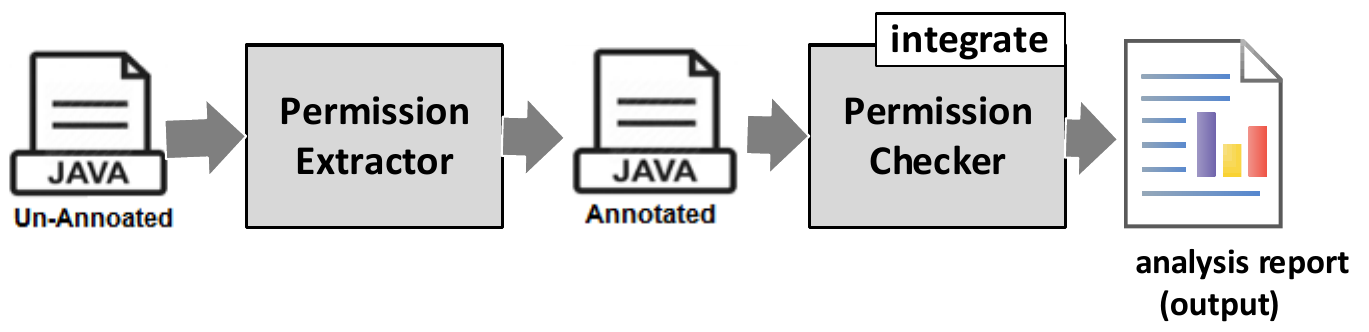}
\caption{A high-level work-flow depiction of the Sip4J framework}
\label{fig:sip4jworkflow}
\end{figure}

We propose a framework called \gap to automatically infer access permissions from a given sequential \java program and subsequently verify the correctness of inferred permissions.
Figure~\ref{fig:sip4jworkflow} presents the working process of \gap, which is mainly made up of two modules: (1) Permission Extractor and (2) Permission Verifier.
We now detail these two modules in Subsection~\ref{sec:method} and Subsection~\ref{perm-checker}, respectively.

\subsection{The Permission Extractor}\label{sec:method}
The permission extractor performs inter-procedural static analysis of an un-annotated \java program based on its Abstract Syntax Tree (AST), the syntactic rules (Section \ref{GC}) and the permission inference rules (Section \ref{TR}). The technique automatically reveals the implicit dependencies present between the code (methods) and the global states (class fields) in a \java program and maps them in the form of access permission rights using graph abstractions. 



The proposed technique includes the steps or phases as shown in Figure~\ref{fig:Figure1b}:
\begin{figure}[H]
\includegraphics[width=\linewidth]{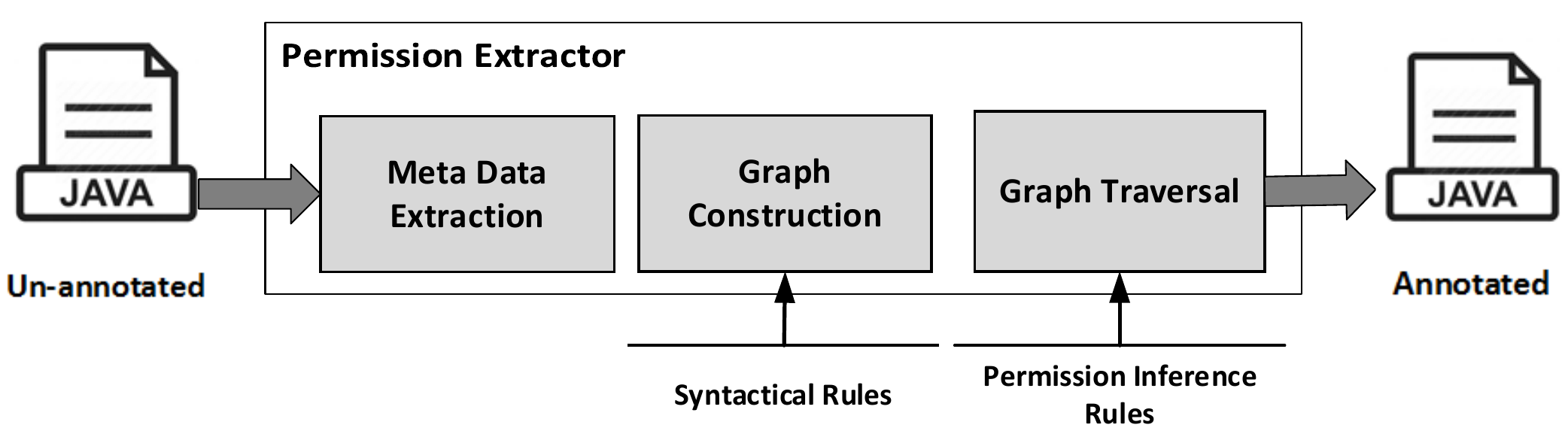}
\caption{The permission extractor phases.}
\label{fig:Figure1b}
\end{figure}
\begin{description}
\item [\textbf{Metadata Extraction.}] It parses the AST of the \java source code to extract and maintain the metadata (dependency) information as data flow, alias flow and context information, for all the shared objects (class fields) accessed in a method (Section \ref{meta-analysis}).
\item [\textbf{Graph Construction.}] For each method, based on the extracted information, it constructs a permission-based graph model, using graph notations and by following the pre-defined syntactic rules (Section~\ref{GC}) that specifies the way to model object's accesses in a graph structure.
\item [\textbf{Graph Traversal.}] It traverses the constructed graph for each method and generates symbolic permissions for the objects referenced in a method using access permission inference rules (cf.\ Section~\ref{TR}).
\end{description}

The approach automatically generates five types of access permissions e.g.,  \code{unique, full, share, pure and immutable}, as \code{pre-} and \code{post-permissions} on individual field of an object (class) at the method level. It further generates an annotated version of the input program with permission contracts following the \plural specifications (\ref{plural-background}) where permission are defined on the receiver object (\code{this}). The pre-permissions are the permissions that caller (client) of a method must provide on the referenced object(s) before invoking a method or alternatively, the permissions that method requires on the referenced objects (fields, parameters etc.) before being executed. The \code{post-permissions} are generated on the referenced object(s) when the method completes its execution.

It is worth mentioning here that, following the \DbC, a method is responsible to return either the consumed (same) permission back to the caller of method e.g., \code{unique} for \code{unique} or generate some restrictive permission such as \code{full} for \code{immutable}, as post-permission to avoid the data integrity problems when permissions are actually used for verification or parallelization purpose. However, in certain cases, the pre- (post) permission on a referenced object could be some special case e.g., the \code{none} permission.

For example, in a \java program, it can happen in three situations  a) if a method (re-)instantiates a (global) referenced object in its local environment, b) if a method itself is the \code{main()} method from where execution of the program starts, c) if a method creates a null reference or un-instantiates a (global) referenced object. For the first two cases, the approach generates \textsf{none} as pre-permission showing the absence of permission which means that the method does not require any permission on the referenced object, to start its execution. In the last case, the approach generates \textsf{none} as post-permission on the reference object

Another special case is when a method reads an object's or class field that is not being accessed by other methods in any way, in this case, the pre- and post-permission for the referenced object would be \code{none}. Moreover, no permission contract would be generated for a method if it does not access any object's (class) field from its global environment or if it manipulates only the local references declared in it.
In the following sections, we will elaborate on each of the three stages of the above mentioned permission inference technique using method \code{maniuplateObjects()} shown in Listing \ref{lst:ua-specprog} at line 56.
\subsubsection{Metadata Extraction}\label{meta-analysis}
To generate access permissions for the shared objects (global states) at the method level, following the access permission semantics (Section \ref{sec:accessperm}),
we need to identify the way (\code{read} or \code{write}) a referenced object\footnote{The term reference variable and reference object has been interchangeably used throughout the paper, to represent a referenced object} is accessed by the current method and, at the same time by its context ("the rest of the world" except the current method). Moreover, we need to identify and track aliases of the referenced objects (if any), to extract correct dependencies and to maintain the integrity of the data during analysis.
For this purpose, the inference approach performs inter-procedural static analysis of an un-annotated \java program based on its Abstract Syntax Tree (AST). It performs data-flow, alias-flow and context analysis of the source code to
extract and maintain the read, write, aliasing information for all the object’s (class) fields
accessed in the current method and their context (access by other methods) information.

The \textbf{data flow and alias flow analysis} of the source code is based on the type of expression encountered in an expression statement such as  \code{<FieldAccess>}, \code{<Assignment>} etc., and the type of referenced variable (object) accessed in each expression, such a class field or a method's local variable (parameter) that is alias of some global reference (class field). We ignore all the local variables and parameters that are not alias of any global reference, this is because manipulating local objects in a method does not affect the access rights of the current and other methods. Further, the context analysis (access by other methods) of a referenced object is based on its data flow and alias flow analysis across other methods. 

For each method in a \java program, the technique extracts the read, write and aliasing information for a referenced object in a method by parsing method's signature and its body in following ways:



\textbf{Method Signature:} 
The extractor parses a method's signature with its name, return type, visibility modifier,and formal parameters.
The formal parameters are mapped with their corresponding argument (aliased) objects by fetching the method invocations of the corresponding method in the program. This information is then used to extract (maintain) the read, write and aliasing information of the actual objects against parameters and to avoid the data inconsistency problems while the same object is being accessed by other methods. For example, in Listing \ref{lst:ua-specprog} for method \code{manipulateObjects()} in Line 56, the technique maps the formal parameters \code{p1} and \code{p2} with their actual objects i.e. \code{w} and \code{z} respectively and then track them in the method body to identify their metadata information.

\textbf{Method Body:}
In parsing a method, the technique parses each expression statement in the AST of a method body.
Each expression is iteratively parsed to distinguish (fetch) the \code{<read-only>} and \code{<read-write>} expressions and these information are recorded accordingly. 
Like parameters, the technique maps all the local references with their global reference (alias) if any, to extract and maintain the data flow and alias flow information of the actual referenced object during parsing.
The handling of read and write expressions in a statement is as follows: 


\begin{itemize}
\setlength{\itemsep}{4pt}
\item[$\bullet$] The \code{<read-only>} expressions are characterized by expression nodes such as  \code{<FieldAccess>}, \code{<QualifiedName>}, \code{<SimpleName>}, \code{<MethodInvocation>} etc., in the AST and parsed to extract all the referenced objects accessed in the expression. This information is maintained in the system as a part of data flow (read access) analysis in the current method.
For example, in Listing \ref{lst:ua-specprog} at line 61, the technique parses \code{p2.data} expression.  The information is maintained as a read access for the referenced object \code{z} in the current method \code{manipulateObjects()}.

\item[$\bullet$] The \code{<read-write>} expressions are characterized by \code{<Assignment>} expressions in the AST. The proposed technique performs flow-sensitive analysis of the source code, where the type of a reference on the left-hand side of an assignment statement is determined based on its right-hand side expression. 

During parsing, the assignment expressions are further categorized as \code{<value-flow>}, \code{<object-creation>}, \code{<address-flow>}, \code{<null-address-flow>}, \code{<self-address-flow>}, etc.,) expressions based on the type of the right-hand side expression and the data type of its referenced variables. 
For example, if the data type of the right-hand side expression is a \code{<Primtive>} type or if it is a \code{<NumberLiteral>} expression, the approach treats the assignment expression as a \code{<value-flow>} expression.
The \code{<object-creation>} expressions are characterized by the presence of \code{<ClassInstanceCreation>}, \code{<ArrayCreation>}, \code{<ArrayInitializer>} expressions on the right-hand side of an assignment expression.
Similarly, If the right-hand side of an assignment statement yields a \code{<ReferenceType>} or a \code{<NullLiteral>}, the expression is categorized as a \code{<null-address-flow>} and a \code{<address-flow>} expression respectively. The extractor recursively parses the right and the left side of an assignment statement to identify the \code{read-only} and \code{read-write} expressions following the expression types and extracts (maintains) the read, write and aliasing information of all the object's (class) fields accessed in each expression.

Let us take Listing~\ref{lst:ua-specprog} again as an example.

The expression \code{y = t;} in Line 59, is categorized as an \code{<address-flow>} expression. 
The technique maps the local reference \code{t} with its actual (global) reference object i.e., \code{x}, as its alias.
This information is maintained as a part of alias flow analysis for reference x.
 
Similarly, the expression \code{x.data = 10;} in Line 60, is treated as a \code{<value-flow>} expression as the right hand side of the expression is a <NumberLiteral> expression. The information is updated as write access for the reference variable \code{x} in the current method. Further, the analysis ensures that this change (write operation) should be propagated to all the aliases of \code{x} i.e., reference \code{y} and \code{w} in this case, to ensure the integrity of data.

\item[$\bullet$] The \code{\textbf{method calls}} in are handled using \code{<MethodInvocation>} and \code{<SuperMethodInvocation>} etc., expressions in the AST. As a part of the modular analysis, the permission inference technique is recursively applied to every callee methods (a non-recursive method call) in the caller method. For this purpose, the current state of the caller method is saved and restored when all of its sub-methods have been parsed. The analysis of the caller method does not been completed until the metadata of all of its sub-methods have been extracted.
For example, in Listing~\ref{lst:ua-specprog} in Line 38, the approach first parses method call expressions \code{printColl()}, \code{isSorted()} and \code{findMax()} in the given order, to extract the read, write and alias information on the object referenced by the parameter \code{coll}. The extracted information constitutes the metadata information for the caller method i.e., \code{computeState()}. The approach then uses this information to generate the necessary permissions for the caller methods.


\item It is worth mention that our technique does not parse the \textbf{recursive method calls} (where a method call itself in its body) expressions. This is because a recursive method call does not change the way the caller method (itself) accesses the referenced objects, reducing the analysis time as well. 
In case of infinite and chained recursion (eg. when a method say \code{foo1()} calls method \code{foo2()} that in-turn calls \code{foo1()} in its body), the analysis terminates successfully. This is because, a) the underlying approach maintains metadata (at least signatures), of each method's say \code{foo1()} before a indirect recursive call say \code{foo2()}, encounters in its body which helps to identify and skip the second level recursive call for the same method \code{foo1()} in the caller method \code{foo2()} and the system continues parsing from the next expression in method \code{foo2()} b)
For super method calls, in the case of inheritance, are handled the same way (parsing its method signature and body) as other non-recursive method calls. 


\item Further, the structure (conditional and dynamic structure) of source program does not affect the permission inference mechanism as extractor parses all expressions encountered in an expression statement based on the type of expression and the type of object referenced in it irrespective of their access location. 
The analysis generates safe (write instead of read) permissions when an object is accessed in multiple expressions
or when the same object is accessed inside dynamic expressions such as \code{switch cases, if-else and loops} etc., by updating the object's meta data across all the expressions in a method.

\item Moreover, handling of the array data structures is the same for single- and multi-dimensional
arrays. This is because, at the moment, the approach does not generate
permissions on the individual elements or dimensions in an array data-structure. It parses
the whole array object like an ordinary instance (class) variable.
\end{itemize}

The \textbf{context analysis} of all the objects accessed in the current method is based on their read and write access by other methods in the program. There can be three possible contexts namely \contextn (no access), \contextr (read-only) or \contextrw (read and write) for a object accessed in the current method. The \contextn is the most restrictive context as in this context, the current method demands exclusive rights (both read and write) on the object; thus reducing the chances to achieve maximum parallelism across methods. The read context (\contextr) is less restrictive than \contextn as it allows some kind of read access to other references thus providing the possibility to achieve more parallelism. \contextrw is the most flexible context as other references would have both read and write access to the referenced object but the possibility of parallelism is limited due to the expected (undesirable) effects, such as data races.

The approach automatically identifies the context information (\code{read, write and none}) for all the shared objects, following their data-flow and alias-flow information in the program. It ensures to extract the safe \context (access by other methods), for the shared objects, by updating their accesses across other methods. For example, in Listing~\ref{lst:ua-specprog}, for method call \code{printColl(coll)} in Line 38, the approach generates \contextrw for reference object \code{array1}, as a part of its context analysis. This is because the array object is being written by other methods such as \code{incrColl()} and \code{tidyupColl()} in the program.

The exceptions to this rule are the objects accessed in \code{<object-creation>} and \code{<null-address-flow>} expressions where current method (un)instantiates an object in its body. In this case, the approach generates \contextn (no-access by other methods) for the reference variable accessed on the left-hand side of an assignment expression, by following the expression type of the right-hand side expression. For example, in Listing~\ref{lst:ua-specprog}, for expression \code{(this.array1 = null;)} in Line 43, the approach generates \contextn as its context information for \code{array1}. This is because the current method un-instantiates the referenced object and it should have exclusive access to create a null-reference (a reference variable that does not refer to any object) and update this information to all of its alias(es).

Similarly, for <object-creation> expression such as \code{array1 = new Integer [10];} in Listing~\ref{lst:ua-specprog} at Line 4, the approach generates \contextn for array object \code{array1}, as the current (constructor) method \code{ArrayCollection()} instantiates a new object and, at this moment, no other method can access it. This information is then used to construct a permission-based graph model of the current method and generate pre- and post-permissions for the referenced objects accessed in it. The type of access permissions generated in each context depends on the way (read or write) the current method accesses the shared object.

It is worth mentioning here that in a \java program execution starts from the \code{main()} method,  therefore execution of the \code{main()} method is independent of any context which means it does not require any (pre-) permission to access the objects from its global environment. The approach ensures to generate \contextn (none) for all the referenced variables accessed in the \code{main()} method. Following, the permission semantics in Section \ref{sec:accessperm}, application of \contextn in-turn generates \textsf{unique} permissions on the referenced object.
\subsubsection{Graph Construction}\label{GC}
\begin{figure}[H]
\centering
\includegraphics[width=0.7\linewidth]{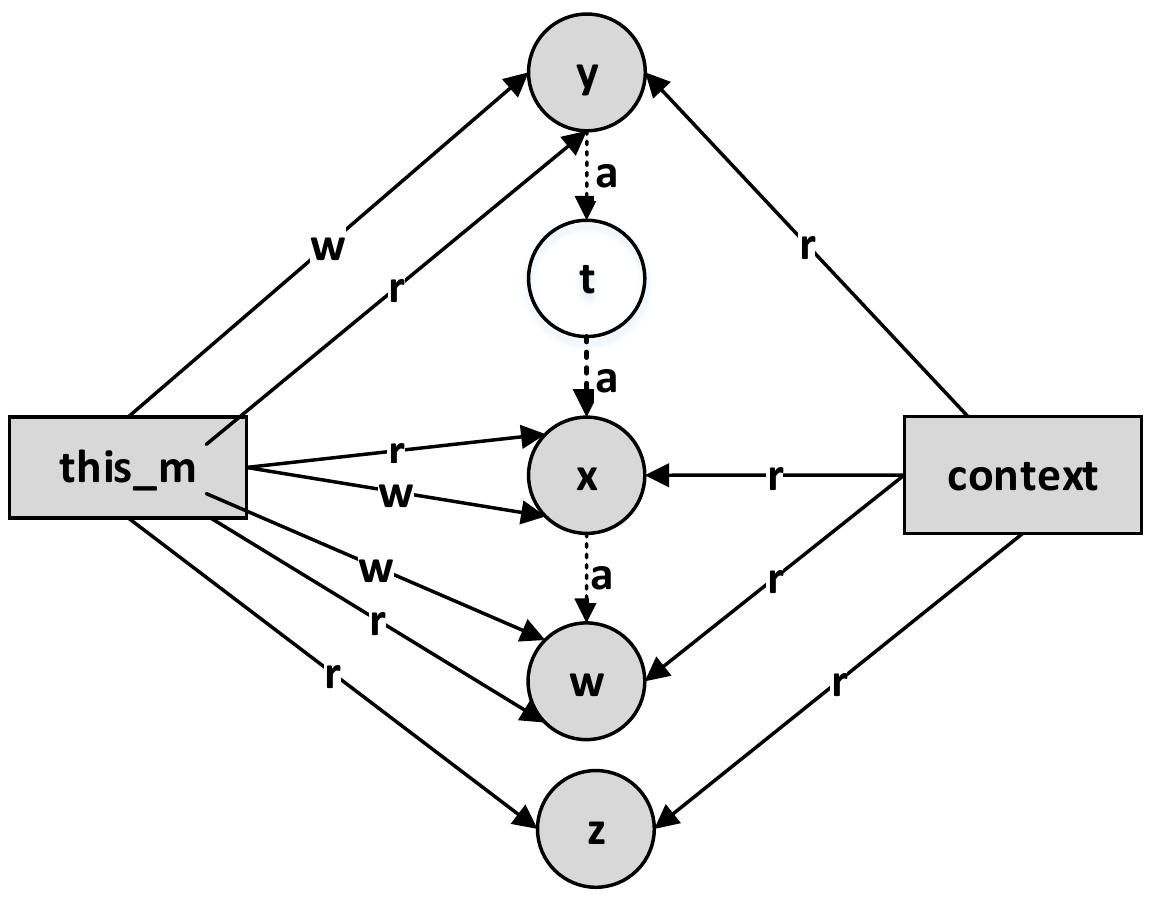}
\caption{The access permission graph model of method maniuplateObjects()}
\label{fig:demo-graph}
\end{figure}

The graph construction phase constructs a permission-based graph model for each method. It maps the extracted data flow, alias flow and context information, for all the (global) referenced objects accessed in a method, using graph abstractions. Figure \ref{fig:demo-graph} shows the permission-base graph model of method \code{manipulateObjects()} generated by the \gap framework. 

\begin{enumerate}[wide, labelwidth=!, labelindent=0pt, label=\Roman*., start=1]
\setlength{\itemsep}{5pt}
\item \textbf{Graph Notations}\label{Gnotations}

To construct a permission-based graph model, we use some special nodes, edges and conventions which are described below:

\begin{description}
\setlength{\itemsep}{4pt}
\item \textbf{Graph Nodes:} The access permission graph is made up of three types of nodes:

\begin{itemize}
\setlength{\itemsep}{3pt}
\item[$\bullet$] A \variable node, depicted as a labelled circle, models a reference object \code{o} accessed by the current and (or) its context. For example, \code{x} in Figure \ref{fig:demo-graph} represents an object accessed in method \code{manipulateObjects()} in Listing \ref{lst:ua-specprog}.

\item[$\bullet$] \foo is an abstraction, a labelled rectangle, that represents the current method accessing the referenced object . For example, method \code{manipulateObjects()} in Listing \ref{lst:ua-specprog} is labelled as \foo in Figure \ref{fig:demo-graph}.

\item[$\bullet$] \context is an abstraction, depicted as a labelled rectangle, that represents the collective access of other methods accessing the same object or the current method’s global environment. 

The nodes \foo and \context have been introduced to make graph construction and traversal process simpler.
\end{itemize}

\item \textbf{Graph Edges:} There are two types of edges that model the way (read, write, and alias), a referenced object \code{o} is accessed in the current method \textbf{(\foo)} and by its context (global environment).

\begin{itemize}
\setlength{\itemsep}{3pt}
\item[$\bullet$] \textbf{read/write edges} are depicted as directed and solid edges labelled as `r' or `w'. For example, in Figure \ref{fig:demo-graph}, there exists a read and a write edge between the current method (\foo) and variable node \code{x}. Similarly, objects \code{x, y, w, z} are read by the client method (\code{main()}) in Listing \ref{lst:sa-specprog} so a read edge with label `r' is drawn from the \context node to all of these \variable nodes.

\item[$\bullet$] \textbf{alias edge} is used to model an alias
of a reference, if any. The alias edge is depicted as a directed and dotted edge labelled with letter `a' between two references. For example, \code{x} is an alias of \code{w} in Figure \ref{fig:demo-graph}.

\end{itemize}
\end{description}

\item \textbf{Modelling Object's Access in a Graph Structure}

We formally specify syntactic rules to support the metadata analysis and to model the object's accesses in a permission-based graph model. The rules are based on the type of expression encounter in an expression statement and the type of reference variables accessed in each expression. For example, in the syntactical rules, the notation \var represents a (global) referenced object i.e., a class field or a parameter that is alias of a class field, the notation \localrefvar represents a method's local variable that is an alias of a global reference and \lvar is a method's local variable other than \localrefvar. The rules are categorized into two types depending on their usage: a) Context rules, b) Statement rules.
The statement rules, for simplicity, are further categorized as method call and non-method call rules.

The rules are designed to follows the style of sequent calculus, as shown in equation \ref{eq:eq1}, with logic connectives and implication $\multimap$ operator, that considers rules as formulas (resources) and enforces their
constructive interpretation to extract and map the object’s accesses in a precise way.

\begin{equation}\label{eq:eq1}
\begin{aligned}\frac{\equfont{<Exp-Statement>}}{\equfont{<Rule-Description>}}\equfont{(\textbf{<Rule-Name>, \var})}
\end{aligned}
\end{equation}
The rule's name (\code{<Rule-Name>, \var{}}),
itself follows the type of referenced object (GR for \var, LR for \localrefvar and L for \lvar) and the type of expression (read, write and address flow) encountered.

Although the rules are self explanatory, we describe some of them, to provide an intuition on mathematical specifying them and to validate the permission inference mechanism through them. 

\end{enumerate}

%
\begin{description}
\setlength{\itemsep}{4pt}
\item[$\bullet$] The \textbf{Context} rules \label{contextR} specify the way to add read and write edges
between \context and  \variable nodes. The context of referenced object is specified as \contextn (no access), \contextr (read access) or \contextrw (read and write access). 
For example, the \code{\textbf{(Context-R, \var)}}
rule specifies that we need to add a read edge from \context to \variable node to show that the object (\var) accessed in the current method is also read by other methods.

\[
{\footnotesize
\centering
\begin{tabular}{c c}
\begin{prooftree}
\,\texttt{\var}\,
\using{\textsf{\textbf{(Context-R, \var)}}}
\justifies
\begin{array}{l}
\,\texttt{addReadEdge(context,\var)}\, \\
\end{array}
\end{prooftree}
\end{tabular}
}\]\\

\item[$\bullet$] The \textbf{Statement} rules describe different ways (read and write) to add edges between 
a \variable node and the current method as well as its context (if any).

For example, the \textbf{(GR-Val-Flow,\var)} rule given below, models the write access of the current method (\foo) on left hand side object (\var) in an assignment statement. This rule states that we should add a write edge from current method \foo to the \var node. It further ensures that this change should be propagated to all the aliases of \var to maintain integrity of data during parsing. Therefore, in the graph, we need to add a write edge from \foo node to all its alias(es) nodes, if any. 
\[{\footnotesize
\begin{tabular}{l l}
\begin{prooftree}
\,\texttt{[Prim\_Type] \var = \varnum{1}|<LITERAL>}\,
\justifies
\begin{array}{l}
\,\texttt{addWriteEdge(\foo,\var),apply(GR-Read-Only,\varnum{1}),}\,\\\\
\,\texttt{($\forall a \in$ aliasOf(\var)$\multimap$(addWriteEdge(\foo,a))))}
\end{array}
\end{prooftree}
\end{tabular}
}\]
All the references on right hand side of the assignment expressions are modeled following the appropriate <read-only> rules given in Appendix \ref{rules}.

Similarly, the \textbf{(LR-Addr-Flow, \var)} rule models an <address-flow> statement of the form \code{\localrefvar = \var}. The rule states that we should add an alias (pointee edge) edge from local reference (\localrefvar) to the global field (\var) and should remove its existing alias edge, if any. The analysis keeps track of the changes in the state of \localrefvar, as a part of alias-flow analysis, that could affect the object referenced by the global reference (\var) and its aliases.
\[{\footnotesize
\begin{tabular}{l l}
\begin{prooftree}
\,\texttt{[<Ref\_Type>] \localrefvar = \var}\,
\justifies
\begin{array}{l}
\,\texttt{($\exists$aliasEdge(\localrefvar,\varnum{1})$\multimap$removeAliasEdge(\localrefvar,\varnum{1})),}\, \\\\
\,\texttt{addAliasEdge(\localrefvar,\var),apply(GR-Read-Only,\var)}\,
\end{array}
\end{prooftree}
\end{tabular}
}\]
\item[$\bullet$] The \textbf{MCall} \label{MR}
rules capture method invocation expressions. The method call rules
specify the way to add read and write edges in the caller method graph as a result of a (sub) method call. The type of edges added in the caller method graph depends on the post access permissions (\code{<post-perm>}) generated by a called method on its referenced object(s) (\var). This is because, when actually parallelizing code based on permissions, the caller
method needs to provide the pre-permissions for all the object’s (class) fields accessed
in the called method, to execute the called method as a part of its own execution.

For example, for a method call \code{\textbf{MCall(<Full>, \var)}} that generates \textsf{full} permission on its reference object (\var), as post permissions, we need to add both a read and write edge from the called method (\foo) to the \variable (\var) node and apply the read context rule on \var,  by following the semantics of \textsf{full} permission.

\end{description}

\[
{
\resizebox{\linewidth}{!}{
\begin{tabular}{l l}
\begin{prooftree}
\,\texttt{MCall([<args>])|super.MCall([<args>])|super([<args>])}\,
\using{\textsf{\textbf{MCall(Full, \var)}}}
\justifies
\begin{array}{l}
\,\texttt{addreadWriteEdge(\foo,\var),apply(Context-R,\var)}\, \\
\end{array}
\end{prooftree}
\end{tabular}}
}\]
Other expressions are handled similarly by following the appropriate syntactic rules. A complete list of syntactic rules for modelling object's accesses in a graph model is given in Appendix \ref{rules}.

\subsubsection{Graph Traversal \& Permission Generation}\label{TR}
The graph traversal phase generates five kinds of symbolic permissions on the referenced objects at the method level by traversing the permission-based graph model of each method and following the access \textbf{permission inference} rules. The type of access permissions generated depends on the type of edges between the current method (\foo) and the \variable nodes and the presence (or absence) of alias edges between \variable nodes. For example, the \textbf{(Full, \var)} permission inference rule
states that:
\begin{itemize}
\item[$\bullet$] There must not be a write edge from \context to \var node; and
\item[$\bullet$] There must exist a read and a write edge from \foo to \var node.
\end{itemize}

\[{\footnotesize
\begin{tabular}{c c}

\begin{prooftree}
\begin{array}{l}
\,\texttt{$\exists$readWriteEdge(\foo,\var) $\wedge$ $\exists$readEdge(context,\var)}\,\\\\
\,\texttt{$\wedge$ $\neg$$\exists$writeEdge(context,\var)}\,
\end{array}
\using{\textsf{(\textbf{Full})}}
\justifies
\begin{array}{l}
\,\texttt{Full(\var)} \,
\end{array}
\end{prooftree}
\end{tabular}
}\]

Listing \ref{lst:sa-specprog} shows the annotated version of the input \java program given in Listing \ref{lst:ua-specprog}, generated by the \gap framework, with access permission contracts at the field level. In Listing \ref{lst:sa-specprog}, Line 35 \& 36 shows the permission contract (as pre- and post-permissions) for the method \code{manipulateObjects()}, by traversing its permission graph model given in Figure \ref{fig:demo-graph} and by following the \textsf{Full} and \textsf{Immutable} permission inference rules.
It is worth noting here that all the parameters and local references, in Figure \ref{fig:demo-graph}, are mapped
with their global references to generate permissions on the actual referenced objects.

Other kinds of access permissions are generated using their corresponding rules. A complete list of access permission inference rules is attached in Appendix \ref{AP-rules}.

\subsection{The Permission Checker}\label{perm-checker}

As the second module of \tool{}, the permission checker aims at verifying the correctness and effectiveness of the inferred specifications. As discussed previously in Section \ref{plural-background}, for this purpose, it extends and integrates \pulse,  a permission-based verification tool that verifies correctness of the \plural specifications, i.e., a \java program annotated with access permission contracts and typestate information, in isolation to program code. 
The permission checker extends the \pulse tool in following three ways.
\begin{description}
\setlength{\itemsep}{5pt}
\item[\textbf{(1) Pre-Processing:}] 
As a part of pre-processing, \gap automatically generates the \pulse translated version of the input program with \plural specifications where permissions are defined at the object (using keyword \ ``\code{this}'') level and that follows the \DbC to generate pre- and post-permissions as a part of method contract. It is worth mentioning here that \pulse does not support overloaded methods, as a part of its analysis, even if the method is provided with different method signature and permission contracts. 

The permission checker makes following changes in the annotated \java program given in Listing \ref{lst:sa-specprog} to generate an input (\plural annotated) program acceptable by the \pulse tool.

\begin{itemize}
\setlength{\itemsep}{4pt}
\item It generates a non-parameterized (default) constructor, even if no explicit constructor is defined in the class, to generate \textsf{unique} permission on the receiver object (\code{this}) (see Listing \ref{lst:a-pluralspecprog} line\ 4), that is later used to perform permission analysis of other methods accessing the same object. 

\item For non-constructor methods, it generates conservative and safe permission for the receiver object (\code{this}) using the permissions generated at the field level. For this purpose, it computes maximum of the pre- and post- permissions of all the individual fields associated with the current instance of object (see Listing \ref{lst:a-pluralspecprog} line\ 15 \& 16).

\item For non-constructor methods, it generates notation \code{<AP>(\#i)} to represent pre- and post-permissions for the referenced parameters (see Listing \ref{lst:a-pluralspecprog} line\ 15 \& 16).

\item For non-constructor methods, it generates permission contracts following the relation $P \mbeq Q$ where pre (P) and post (Q) conditions should be same (see Listing \ref{lst:a-pluralspecprog} line\ 7 \& 8).

\item It automatically add other required annotations as a part of annotated program such as:
     \begin{itemize}
     \setlength{\itemsep}{4pt}
     \item An import statements to support \plural annotations (Listing \ref{lst:a-pluralspecprog}, line 1) as a part of program.
     \item It automatically add typestate `\code{alive}' using \code{@States} statement at the class level (Listing \ref{lst:a-pluralspecprog}, line 2).
     \item Typestate as a part of pre- and post- permission (Listing \ref{lst:a-pluralspecprog}, line 3).
     \item The annotation \code{ENDOFCLASS} at the end of each class (Listing \ref{lst:a-pluralspecprog}, line 10 \& 19).
     \end{itemize}
\end{itemize}
Listing \ref{lst:a-pluralspecprog} shows the \plural annotated version of methods \code{tidyupColl()}, \code{manipulateObjects()} and the default constructors in \code{ArrayCollection} and \code{ObjectClass} class.

The annotations generated for the \pulse translated version of the input program also shows the minimum annotation overhead imposed by the existing permission-based verification approaches such as \plural and \pulse to verify program behavior.
\begin{figure}[H]
\begin{lstlisting}[morekeywords={none,full,unique,ensures,requires,alive,ALIVE,alive,share,immutable,pure,main},
label=lst:a-pluralspecprog,caption={The Plural annotated version of some of the methods given in Listing \ref{lst:sa-specprog}.}]
import edu.cmu.cs.plural.annot.*;
@States({@State(name = "alive")})
class ArrayCollection{
@Perm(ensures="unique(this) in alive")
ArrayCollection() {   }
@Perm(requires="unique(this) in alive * unique(#0) in alive",
ensures="unique(this) in alive * unique(#0) in alive")
public void tidyupColls(Integer[] coll) { } 
}
ENDOFCLASS
@States({@State(name = "alive")})
class ObjectClass{
@Perm(ensures="unique(this) in alive")
ObjectClass() {   }
@Perm(requires="full(this) in alive * full(#0) in alive * pure(#1) in alive",
    ensures="full(this) in alive * full(#0) in alive * pure(#1) in alive")
void manipulateObjects(Client p1, Client p2) { }
}
ENDOFCLASS

\end{lstlisting}
\end{figure}

\item[\textbf{(2) Extended Concurrency Analysis:}]
The permission checking approach extends the \pulse analysis to perform a comprehensive concurrency analysis of sequential \java programs based on inferred permissions. 

In \pulse, two methods are considered parallel if both require \code{pure}(read access) as pre-permission on the referenced object. The non-parallel behavior of the two methods is determined based on \code{unique} and \code{full} permissions which means the methods that require read-write (\code{unique} or \code{full}) access cannot be parallelized with each other.
However, the concurrency analysis in the \pulse tool is limited in two ways a) it does not consider all the possible side effects comprising \code{full} and \code{pure} permissions, b) it does not support \code{immutable} and \code{share} permissions, as a part of its concurrency analysis, and consequently, the model-checker is not able to consider the method's side effects for following pairs of (pre-)permission contracts between two methods e.g., \code{\{(full, pure), (pure, full), (share, pure), (share, share)}\}, and reports them to be concurrent. However, methods with these specifications, as a part of the method contract, if allowed to execute in parallel can cause data races of the form \code{<write-read>}, \code{<read-write>} or \code{<write-write>}. 

Table \ref{exp-concurrent-pair} shows the expected method-pair concurrency analysis that should be performed for a sequential program to void data races when the program is actually parallelized based on access permissions. The symbol \textcolor{blue}{$\parallel$} indicates the parallel execution of the two methods, whereas the symbol \textcolor[rgb]{1.00,0.00,0.00}{$\nparallel$} shows the fact that the specified methods should be executed in parallel with each other.

\begin{table}[htb]
\centering
\caption{Expected method-pair concurrency analysis.}\label{exp-concurrent-pair}
{\renewcommand{\arraystretch}{1.0}
\begin{tabular}{lrcccccc}
\multicolumn{1}{l}{} && \multicolumn{5}{c}{APs(m2)} \\ \cline{3-7}
\multicolumn{1}{l}{} & &  &  &  & &   \\
\multicolumn{1}{l}{} & & unique & full & share &immutable & pure\\ \hline
\multirow{5}{*}{\begin{sideways}APs(m1)\end{sideways}}
& \multicolumn{1}{|r}{unique} & \textcolor{red}{$\nparallel$} &\textcolor{red}{$\nparallel$} & \textcolor{red}{$\nparallel$}& \textcolor{red}{$\nparallel$} & \textcolor{red}{$\nparallel$}       \\ \cline{2-7}
& \multicolumn{1}{|r}{full} & \textcolor{red}{$\nparallel$} & \textcolor{red}{$\nparallel$}  & \textcolor{red}{$\nparallel$}  & \textcolor{red}{$\nparallel$}  &  \textcolor{red}{$\nparallel$}   \\ \cline{2-7}
& \multicolumn{1}{|r}{share} &  \textcolor{red}{$\nparallel$} & \textcolor{red}{$\nparallel$}  & \textcolor{red}{$\nparallel$}  & \textcolor{red}{$\nparallel$} & \textcolor{red}{$\nparallel$}       \\ \cline{2-7}
& \multicolumn{1}{|r}{immutable} & \textcolor{red}{$\nparallel$}  & \textcolor{red}{$\nparallel$}  &  \textcolor{red}{$\nparallel$} &  \textcolor{blue}{$\parallel$} &  \textcolor{blue}{$\parallel$}    \\ \cline{2-7}
& \multicolumn{1}{|r}{pure} & \textcolor{red}{$\nparallel$} & \textcolor{red}{$\nparallel$} & \textcolor{red}{$\nparallel$}  & \textcolor{blue}{$\parallel$}  & \textcolor{blue}{$\parallel$} \\     \bottomrule
\end{tabular}
}
\end{table}

Further, in the \pulse tool, the model-checker does not perform concurrency analysis of the program using pre-permission contracts of the form \code{(immutable,immutable)} and \code{(pure,immutable)} between two methods. Table \ref{pulse-concurrent-pair} shows the method-pair concurrency analysis performed by the \pulse tool where the symbol \textcolor{blue}{$\checkmark$} indicates the options where \pulse identifies the method's side effects correctly, whereas the symbol \textcolor[rgb]{1.00,0.00,0.00}{$?$} shows the option where either the \pulse too does not support the permission annotations or performs incorrect analysis. 

\begin{table}[H]
\centering
\caption{Method-pair concurrency analysis in Pulse.}\label{pulse-concurrent-pair}
{\renewcommand{\arraystretch}{1.0}
\begin{tabular}{lrcccccc}
\multicolumn{1}{l}{} && \multicolumn{5}{c}{APs(m2)} \\ \cline{3-7}
\multicolumn{1}{l}{} & &  &  &  & &   \\
\multicolumn{1}{l}{} & & unique & full & share &immutable & pure\\ \hline
\multirow{5}{*}{\begin{sideways}APs(m1)\end{sideways}}
& \multicolumn{1}{|r}{unique} & \textcolor{blue}{$\checkmark$} &\textcolor{blue}{$\checkmark$} & \textcolor{red}{$?$}& \textcolor{red}{$?$} & \textcolor{blue}{$\checkmark$}       \\ \cline{2-7}
& \multicolumn{1}{|r}{full} & \textcolor{blue}{$\checkmark$} & \textcolor{blue}{$\checkmark$}  & \textcolor{red}{$?$}  & \textcolor{red}{$?$}  &  \textcolor{red}{$?$}   \\ \cline{2-7}
& \multicolumn{1}{|r}{share} &  \textcolor{red}{$?$} & \textcolor{red}{$?$}  & \textcolor{red}{$?$}  & \textcolor{red}{$?$} & \textcolor{red}{$?$}       \\ \cline{2-7}
& \multicolumn{1}{|r}{immutable} & \textcolor{red}{\text{?}}  & \textcolor{red}{$?$}  &  \textcolor{red}{$?$} &  \textcolor{red}{$?$} &  \textcolor{red}{$?$}    \\ \cline{2-7}
& \multicolumn{1}{|r}{pure} & \textcolor{blue}{$\checkmark$} & \textcolor{red}{$?$} & \textcolor{red}{$?$}  & \textcolor{red}{$?$}  & \textcolor{blue}{$\checkmark$} \\     \bottomrule
\end{tabular}
}
\end{table}

The \code{immutable} permission being the safe permission, if applicable, can support maximum parallelism between methods without the fear of data races, and \code{share} permission being the most flexible access can create side effects thereby, data races. Therefore, these specifications should be considered, as a part of the concurrency analysis, for sequential programs to parallelize their execution without the fear of data races.

\begin{itemize}[leftmargin=0cm]
\setlength{\itemsep}{4pt}
\item The \gap framework extends the \pulse concurrency analysis, by defining the new discrete state semantics for \code{share} and \code{immutable} permission and by updating the existing discrete state model of the input specifications (write tokens) in case of \code{full} and \code{pure} permissions. The objective was to compute the potential for concurrency in a sequential program, by considering all possible side effects, based on five types of access permissions.

Table \ref{concurrent-pair} shows an adjacency matrix showing methods (pair) concurrency analysis after extending the \pulse analysis. All the blue \textcolor{blue}{$\checkmark$} symbols show the concurrency analysis that the \pulse tool performs based on the permission compatibility and side effects analysis, and all the red \textcolor{red}{$\checkmark$} symbols show the method pair concurrency analysis extended by our approach. The extended concurrency analysis can be used to parallelize execution of \java programs, to the extent permitted by the inferred dependencies, without the fear of data races.

\begin{table}[H]
\centering
\caption{Extended method-pair concurrency analysis.}\label{concurrent-pair}
{\renewcommand{\arraystretch}{1.0}
\begin{tabular}{lrcccccc}
\multicolumn{1}{l}{} && \multicolumn{5}{c}{APs(m2)} \\ \cline{3-7}
\multicolumn{1}{l}{} & & unique & full & share &immutable & pure\\ \hline
\multirow{5}{*}{\begin{sideways}APs(m1)\end{sideways}}
& \multicolumn{1}{|r}{unique} & \textcolor{blue}{$\checkmark$} &\textcolor{blue}{$\checkmark$} & \textcolor{red}{$\checkmark$}& \textcolor{red}{$\checkmark$} & \textcolor{blue}{$\checkmark$}       \\ \cline{2-7}
& \multicolumn{1}{|r}{full} & \textcolor{blue}{$\checkmark$} & \textcolor{blue}{$\checkmark$}  & \textcolor{red}{$\checkmark$}  & \textcolor{red}{$\checkmark$}  &  \textcolor{red}{$\checkmark$}   \\ \cline{2-7}
& \multicolumn{1}{|r}{share} &  \textcolor{red}{$\checkmark$} & \textcolor{red}{$\checkmark$}  & \textcolor{red}{$\checkmark$}  & \textcolor{red}{$\checkmark$} & \textcolor{red}{$\checkmark$}       \\ \cline{2-7}
& \multicolumn{1}{|r}{immutable} & \textcolor{red}{$\checkmark$}  & \textcolor{red}{$\checkmark$}  &  \textcolor{red}{$\checkmark$} &  \textcolor{red}{$\checkmark$} &  \textcolor{red}{$\checkmark$}    \\ \cline{2-7}
& \multicolumn{1}{|r}{pure} & \textcolor{blue}{$\checkmark$} & \textcolor{red}{$\checkmark$} & \textcolor{red}{$\checkmark$}  & \textcolor{red}{$\checkmark$}  & \textcolor{blue}{$\checkmark$} \\     \bottomrule
\end{tabular}
}
\end{table}

\item It computes the percentage of method pairs that can be parallelised (including the method with itself) over a state space of (all possible method pairs) method pairs in a given program. 
For each class, it calculates the number of concurrent (method) pairs following the binomial coefficient formula given below, choosing two (02) methods from \code{n} methods.
\end{itemize}

\[
    \binom{n}{2} + n = \frac{n!}{k!(n-k)!} + n
\]

where \code{n} is the total number of methods having permission contracts.
Figure \ref{fig:concur_analysis} shows the extended method-pair concurrency analysis of the \pulse translated version of \code{ArrayCollection} class performed by the \gap framework. 
\begin{figure}[htb]
\begin{center}
\begin{subfigure}[t]{0.5\textwidth}
\centering
\includegraphics[scale=.9]{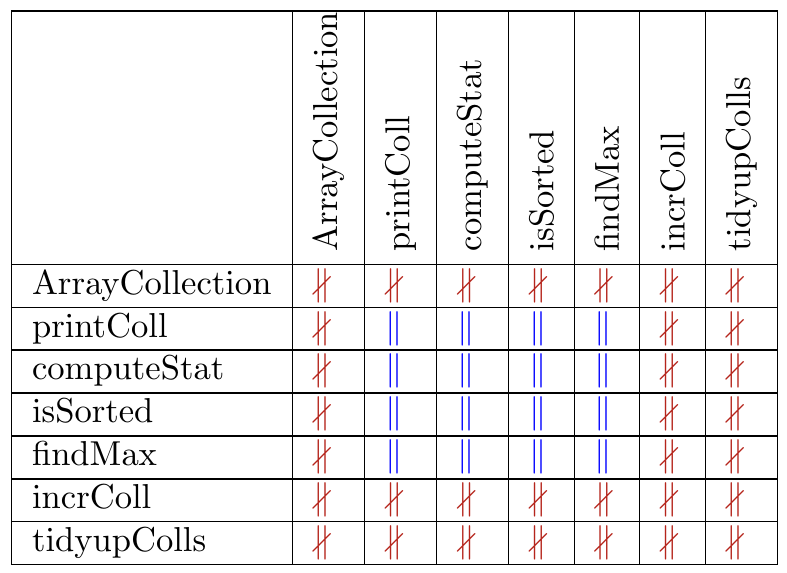}
\caption{Concurrency matrix for all the methods in the ArrayCollection class.}
\label{fig:concur_analysis}
\end{subfigure}

\begin{subfigure}[t]{0.5\textwidth}
\centering
\includegraphics[scale=.9]{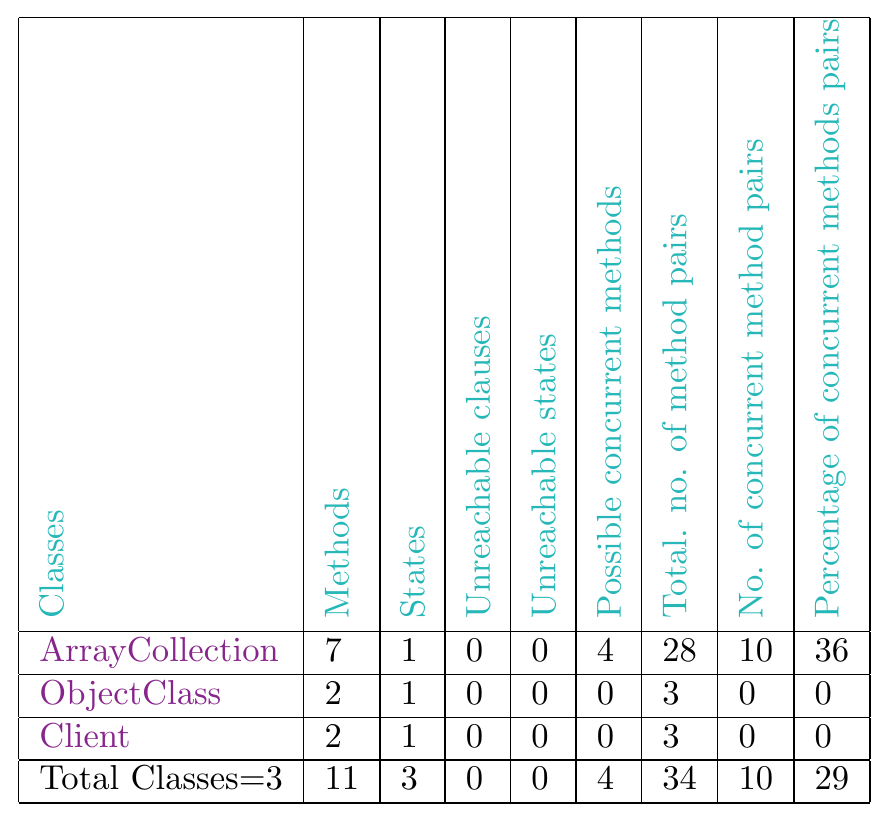}
\caption{Summary of the correctness and concurrency analysis for Plural annotated version of the example program given in Listing \ref{lst:sa-specprog}.}
\label{fig:example-summary}
\end{subfigure}
\end{center}
\end{figure}

The analysis shows that 4 out of 7 (57\%) methods could potentially be executed with at least one other method at the class level. Moreover, the \gap concurrency (method pair) analysis shows that, 10 pairs (36\%) of methods from a state space of total $\tbinom{7}{2} + 7 = 28$ pairs in this class can be run in parallel.

\item[\textbf{(3) Report Generation:}]
\gap extends the \pulse generated (Pdf) report in two ways a) by automatically adding the \plural annotated \java program as a part of generated report and b) by incorporating the results of extended (method-pair) concurrency analysis of the input program in the generated report. Figure \ref{fig:example-summary} shows a summary of the correctness and concurrency analysis of \code{ArrayCollection} class in the generated report.
The objective was to provide readers a quick, explicit and an integrated view of the underlying program with the inferred permissions (method-level dependencies) along with its correctness, code reachability, and concurrency results at one place.
\end{description}

\section{Evaluation}\label{evaluation}
We implemented permission inference framework, \gap, along with its integration with \pulse as a \java Eclipse Plugin. We performed the empirical evaluation of the \gap framework on computationally intensive \java applications from four benchmarks suites, \jomp (jomp)~\cite{jomp}, \aem{}\footnote{\url{https://github.com/AEminium/AeminiumBenchmarks/tree/master/src/aeminium/runtime/benchmarks/}.} and \plaid{}\footnote{\url{https://github.com/plaidgroup/plaid-lang}.} and Crystal\footnote{\url{https://code.google.com/archive/p/crystalsaf/}}, together with \pulse{} itself.

The evaluation of the permission inference framework and the inferred specifications is based on the following research questions and the experiments.
\begin{description}
\item[\textbf{RQ1:}] How can we validate the correctness of the inferred specifications? For this, we automatically perform the \textbf{correctness} analysis of the inferred specifications using \pulse and by manually generating the specifications from the source code and comparing them with the inferred one (Section \ref{correctnessAnalysis}).
\item[\textbf{RQ2:}] How can we demonstrate the effectiveness of the inferred specifications to enable concurrency? We automatically perform the \textbf{concurrency analysis} of the inferred specifications (Section \ref{correctnessAnalysis}) by extending the \pulse concurrency analysis and by computing the number of concurrent method pairs in a sequential program.
\item[\textbf{RQ3:}] How effective is the permission inference technique itself? We automatically compute the \textbf{annotation overhead} (effort saved) by calculating the number of annotations produced by the \gap.
\item[\textbf{RQ4:}] How efficient is the permission inference technique? We compute \textbf{execution time} of the analysis to automatically generate specifications.
\end{description}

\subsection*{Experimental Setup}
All experiments were performed on MacBook Pro, Intel Core i7, (2.3GHz) processor (4 physical cores) and 16GB of RAM. The development environment includes Eclipse IDE 3.7.2, JDK 1.7 and Antlr compiler 3.3 and TexLive 2015.

\subsection*{Datasets}
The dataset for the evaluation consists of benchmark programs and realistic \java applications widely used in the research community \citep{jomp,aldrich2011plaid,CatanoASA14,aldrich2012plaid,Aeminium2014,fonseca2016automatic} to evaluate the permission-based program verification approaches and to gain performance improvements in automatic parallelization approaches. A brief characteristic of the benchmark programs is given below.

\begin{description}
\item[\jomp Benchmark] is a \jomp benchmark suit\footnote{\url{https://www.epcc.ed.ac.uk/research/computing/performance-characterisation-and-benchmarking/java-grande-benchmark-suite}}, consists of scientific and large scale computation intensive applications such as \code{monetcarlo}, \code{moldyn} etc., aimed at testing performance improvement through \java execution environments.
\item[\aem and \plaid Benchmark:]  All the programs in the \aem benchmark consist of data and computationally intensive applications that mostly solve problems using \code{Arrays} and \code{Collection} data structures in \java such as de Columbian Health Care System\cite{healthsystem} and gaknapsack program. The applications were taken from different sources such as HPCC (High Performance Cluster Computing)\footnote{\url{https://icl.utk.edu/hpcc/}}, BOT (Barcelona OpenMP Tasks Suite)\citep{BOT} and \java ForkJoin framework \citep{JFJF}. All the applications in this benchmark are part of \aem \footnote{\url{http://aeminium.dei.uc.pt/index.php/AEminium}} project, a by-default concurrent programming paradigm, that has been used to evaluate the performance of a single-threaded program on multi-core processors, by parallelizing its execution based on access permissions.

The data-set common in \aem and \plaid benchmark consists of programs implementing recursive data structures and commonly-used divide-and-conquer (computational intensive) algorithms such as Fast-Fourier Transform (FFT), Integral, ShellSort, Fibonacci, and a Webserver applications. 

\item[\pulse] is a permission-based model checking tool, implemented as a \java Eclipse Plugin. The tool was developed, as a part of the \aem project 
, to automatically perform permission checking and verify program behavior based on the specifications. The motivation behind using the \pulse tool itself as a case study was to evaluate the permission inference technique on a real-time \java application of average size, in this case, its 7k plus SLOC. \pulse is an object-oriented program that implements rich \java constructs such as classes, inheritance, method overloading, regular expressions, and it extensively uses \java APIs, especially the multi-level linked lists to perform static analysis of the source code with the input specifications, to generate its abstract state-machine model that is then verified by a symbolic model checker.

\item[Crystal]
is a static analysis framework built as an Eclipse Plugin in \java, developed at Carnegie Mellon University for teaching and research purposes. The static analyzer, \crystalsaf, is an object-oriented \java application, developed as a part of the \plural \footnote{\url{https://code.google.com/archive/p/pluralism/}} project. \plural is a sound modular protocol checking tool that employs access permissions to allow a flexible aliasing control mechanism to verify program behavior. Crystal in \plural performs branch-sensitive data-flow and exceptional control-flow analysis of the \java source code. The data-flow analysis is based on a ''worklist" algorithm. The analysis extracts different information at different program levels to perform program verification based on the input permission specifications.

The motivation behind using the Crystal application is to evaluate the permission inference mechanism on a realistic big case study, in this case, 17k plus lines of the source code, that follows object-oriented concepts such as object encapsulation, object composition, object aggregation, etc. Moreover, it implements rich \javasl constructs such as classes, anonymous classes, inner classes, enumeration classes, abstract classes, method overloading, method overriding, multiple inheritance through interfaces, generics and exceptional handling. 

\end{description}

A brief statistics of the benchmark programs are given in Table \ref{table:benchmarks}. 

\renewcommand{\thefootnote}{\fnsymbol{footnote}}

\begin{table}[htb]
\begin{center}
\caption{A brief statistics of the benchmark programs.}
\label{table:benchmarks}
\begin{threeparttable}
\small
\begin{tabular}{p{50pt}p{50pt}ccc}
\hline\noalign{\smallskip}
\multicolumn{5}{c}{Statistics}\\
\cmidrule(lr){1-5}
\textbf{Benchmark} & \textbf{Program} & \textbf{SLOC}\tnote{$\star$}& \textbf{Classes}  & \textbf{Methods}\\
\noalign{\smallskip}\hline\noalign{\smallskip}
Plural&crystal&17,512&212\tnote{$\star\star$}&1,975\\
\midrule
Pulse&pulse&7,671&40\tnote{$\diamond$}&461\\
\midrule
\multirow{10}{*}{\parbox{50pt}{\textbf{jomp}}}
 &montecarlo&1,370&18&196\\
 &euler&1,080&7&51\\
 &search&666&7&50\\
 &moldyn&608&7&43\\
 &lufact&549&5&42\\
 &crypt&488&5&40\\
 &series&359&5&37\\
 &sor&354&5&34\\
 &sparsematmult&327&4&33\\
\midrule
\multirow{4}{*}{\parbox{48pt}{\textbf{\aem}}}
&gaknapsack&437&6&21\\
&blacksholes&232&6&50\\
&health&232&6&18\\
\midrule
\multirow{5}{*}{\parbox{48pt}{\textbf{\plaid}\tnote{$\dagger$}}}
&webserver&143&3&12\\
&fft&91&4&11\\
&quicksort&66&3&9\\
&shellsort&58&2&7\\
&integral&40&1&5\\
&fibonacci&22&1&4\\
\midrule
\textbf{Example}&ArrayCollection&71&3&12\\
\midrule
\multicolumn{2}{l}{\textbf{Total}}&\textbf{32,376}&\textbf{350}&\textbf{3,111}\\
\noalign{\smallskip}\hline
\end{tabular}
\begin{tablenotes}
\item[$\star$]{\scriptsize{SLOC computed using SLOCOUNT tool at \url{http://dwheeler.com/slocount}}} 
\item[$\star\star$]{\scriptsize{It includes three inner classes.}} 
\end{tablenotes}
\end{threeparttable}
\end{center}
\end{table}

\subsection{Correctness analysis of the inferred specifications}\label{correctnessAnalysis}
The \gap framework automatically checks the correctness of inferred specifications using the method (un) satisfiability analysis in \pulse (previously explained in Section \ref{plural-background}). A method is satisfiable if all of its pre-conditions are met. 
In other words, each method obtains its required permission. The presence of the unsatisfiable method is due to the method's unsatisfiable pre-conditions which either indicates an error in the inferred specifications or the missing specifications. Moreover, we cross check the correctness of the inferred specifications for which we a) manually generate the specifications, for all the benchmark programs, by looking at source code b) compare the manually generated annotations with the specifications inferred by the tool, and the \pulse correctness analysis. Moreover, we compute the number of safe approximations made by the inference technique to generate safe permissions (e.g., \code{Full} instead of \code{Pure} or \code{Immutable}), in some scenerios (explained in \ref{threats}) on the referenced objects.

Table \ref{table:correctness} shows the results of the automatic and manual correctness analysis of all the benchmark programs.

\subsubsection{Pulse correctness analysis}\label{satisfiability}
The column \textbf{$\#$ Satisfiable(M)} in table \ref{table:correctness} shows the number of methods determined by the \pulse to be satisfiable, whereas column \textbf{$\#$ Un\_Satisfiable(M)} shows the number of methods \pulse determines to be unsatisfied (unreachable) in a particular program. In other words, total methods is a sum of satisfiable and unsatisfiable methods for each program. 
The results confirms that \gap successfully infers satisfiable (required or enough) permissions, without any specification error, for all methods in all the benchmark programs with three exceptions: montecarlo, \pulse and \crystalsaf.

For monetcarlo, 11 of the 196 methods have been determined to be un-satisfiable. Upon manual analysis of the montecarlo source code, we noticed that the unsatisfiability is due to the fact that \pulse does not support overloaded methods (constructors), and that all these 11 methods are overloaded methods. For the \crystalsaf, the analysis shows unsatisfiability for 581 out of 1,975 methods. Again, this is due to the presence of overloaded methods in \crystalsaf. It is worth mentioning here that \pulse correctness and concurrency analysis is based on satisfiable methods. Therefore, we excluded overloaded methods to perform the correctness and concurrency analysis of the inferred specifications through \pulse and eventually removed them from the generated report. 
However, we manually analyzed the overloaded methods with their inferred specifications and found them to be satisfiable.
\begin{table}[htb]
\begin{center}
\caption{Correctness analysis of the inferred specifications.}
\label{table:correctness}
\begin{threeparttable}
\small
\begin{tabular}{p{30pt}ccc}
\hline\noalign{\smallskip}
\multicolumn{1}{c}{ }&\multicolumn{3}{c}{Correctness analysis}\\
\cmidrule(lr){2-4}
\textbf{Program}&\textbf{$\#$ Satisfiable(M)} & \textbf{$\#$ Un\_satisfiable(M)} & \textbf{$\#$ Safe\_approx(M)}\\
\noalign{\smallskip}\hline\noalign{\smallskip}
crystal&1,394&581\tnote{$\ast$}&288\tnote{$\ast\ast$}\\
\midrule
pulse&451&10\tnote{$\ast$}&13\\
\midrule
montecarlo&185&11\tnote{$\ast$}&10\\
euler&51&0&15\\
search&50&0&10\\
moldyn&43&0&10\\
lufact&42&0&10\\
crypt&40&0&10\\
series&37&0&10\\
sor&34&0&11\\
sparsematmult&33&0&11\\
\midrule
blacksholes&50&0&0\\
gaknapsack&21&0&0\\
health&18&0&0\\
\midrule
webserver&12&0&0\\
fft&11&0&0\\
quicksort&9&0&0\\
shellsort&7&0&0\\
integral&5&0&0\\
fibonacci&4&0&0\\
\midrule
ArrayCollection&12&0&0\\
\midrule
\textbf{Total}&\textbf{2,509}&\textbf{602}&\textbf{398}\\
\noalign{\smallskip}\hline
\end{tabular}
\begin{tablenotes}
\item[$\ast$]{\scriptsize{It shows the number of overloaded methods (constructors) in the program.}}
\item[$\ast\ast$]{\scriptsize{It includes 83 safe approximations for generics (parameterized) methods.}}
\end{tablenotes}
\end{threeparttable}
\end{center}
\end{table}

\subsubsection{Manual correctness analysis}\label{quality}
In Table \ref{table:correctness}, column \textbf{$\#$ Safe\_approx(M)} shows the number of safe approximations made by \gap for the \pulse translated version of each benchmark program for \code{M} methods.  

A non-zero number indicates the situation where \gap does not produce optimal solution and this happens in certain situations. One such situation is when an object is accessed in a library method call for which we don't have definition of the method and analysis generates safe (write instead of read) permissions on the referenced objects to ensure integrity of the data (the details are given in Section \ref{threats}). For example, in \jomp, all the programs use class libraries (API) i.e, \code{JGFInstrument} and \code{JGFTimer}, from Java Grande Framework with ten (10) library methods called in each program, for which \gap generates safe permissions. In the case of Crystal, the count is 288 as the application is heavily dependant on \java class libraries. It also includes 83 safe approximations for the generics (parameterized) methods, as  the permission inference analysis does not support generics in the \javasl, for the moment, and for which we generate safe permissions.

In total, \gap made 398 safe approximations (permission annotations) for a total of 3,111 methods, for the \pulse translated version of 20 benchmark programs, that is 4\% of the total annotations (10,157) generated by the \gap framework. However, we observed that generating safe permissions does not affect the integrity of the specifications and the program itself when actually used for verification or parallelization purpose, and it does not invalidate the effectiveness of our technique in automatically generating correct specifications from the source program.

\subsection{Concurrency analysis of the inferred specifications}\label{concurrencyAnalysis}
As a part of concurrency analysis, \gap identifies the number of immutable methods that can potentially be executed with each other and the method that should always run sequentially based on access permission contracts, by integrating and extending the \pulse concurrency analysis for five types of inferred permissions as explained in Section \ref{perm-checker}.

Table \ref{table:concurrency} shows the concurrency analysis of all the benchmark programs. The results of the analysis in Table \ref{table:concurrency} shows a lower bound on the number of concurrent method, as discussed previously we exclude overloaded methods from the \pulse translated version of each program. In Table \ref{table:concurrency}, column \textbf{Concur(M) (\%)} shows the overall percentage of methods that could be parallelized with at least one other method (including itself) for \code{M} methods in a program whereas \textbf{Concur(MP) (\%)} is the percentage of total number of method pairs that can be parallelized (including the method with itself) over all possible pairs of methods for \code{M} methods.

For the monetcarlo, the \textbf{Concur(M)} ratio is 49\% and the \textbf{Concur(MP)} ratio is 20\% for 185 (satisfiable) methods in 18 classes. For Pulse, the ratio is 67\% with 299 methods out of 451 method that \pulse reports to be satisfiable.
We observed that the exclusion of overloaded constructors does not affect the potential for concurrency as constructors cannot be parallelized with any other method during program execution. In summary, the permission-based concurrency analysis in terms of number of concurrent methods (Concur(MP)\%) vary from 35 to 65\% for the \jomp benchmark programs, 5 to 66\% for \aem and \plaid benchmark and 67\% for \pulse and \crystalsaf. Overall, the results show considerable potentials in our inferred specifications to enable concurrency in sequential programs, which is one of the motivations of our work.

\begin{table}[htb]
\begin{center}
\caption{Concurrency analysis of the inferred specifications}
\label{table:concurrency}
\begin{threeparttable}
\small
\begin{tabular}{p{45pt}ccc}
\hline\noalign{\smallskip}
\textbf{Program} && \textbf{Concur(M) (\%)} & \textbf{Concur(MP)\%}\\
\noalign{\smallskip}\hline\noalign{\smallskip}
crystal&&944 (67\%)\tnote{$\mu$}&52\%\\
\midrule
pulse&&299 (67\%)\tnote{$\mu$}&34\%\\
\midrule
montecarlo&&92 (49\%)\tnote{$\mu$}&20\%\\
euler&&18 (35\%)&8\%\\
search&&18 (36\%)&9\%\\
moldyn&&16 (37\%)&10\%\\
lufact&&25 (59\%)&17\%\\
crypt&&25(62\%)&17\%\\
series&&21 (56\%)&15\%\\
sor&&16 (47\%)&11\%\\
sparsematmult&&16 (48\%)&10\%\\
\midrule
blacksholes&&9 (18\%)&6\%\\
gaknapsack&&4 (19\%)&6\%\\
health&&1 (5\%)&1\%\\
\midrule
webserver&&8 (66\%)&55\%\\
fft&&5 (45\%)&36\%\\
quicksort&&2 (22\%)&11\%\\
shellsort&&2 (28\%)&17\%\\
integral&&3 (60\%)&20\%\\
fibonacci&&2 (50\%)&30\%\\
\midrule
ArrayCollection&&5 (41\%)&33\%\\
\midrule
\textbf{Total}&&\textbf{1,531 (61\%)}&-\\
\noalign{\smallskip}\hline
\end{tabular}
\begin{tablenotes}
\item[$\mu$]{\scriptsize{It excludes concurrency analysis of the overloaded methods.}}
\end{tablenotes}
\end{threeparttable}
\end{center}
\end{table}

\subsection{Annotation overhead analysis}\label{sec:overhead}
We measure the annotation overhead as a way to quantify the manual (annotation) effort by measuring the (1) the amount of annotations generated by \gap on individual field of the referenced object, (2) the number of annotated lines generated by \gap for the \pulse translated version of the program and (3) the number of individual annotations for the \pulse translated version.
To compute annotation overhead, we use number of methods \textbf{M}, as the basis for evaluation as the permissions are generated at the method level.

\begin{description}
\item[\textbf{$\#$ LOC(M)$_{P}$}] in Table \ref{table:annotation} shows the lines of contracts (permission annotated lines) generated for the \pulse translated version of \textbf{M} methods in a program. It counts one line (a permission contract N) for each method with a `\texttt{requires}' and an `\texttt{ensures}' clause, one line for each class to define typestate information at the class level, and one line to import the package that supports \plural annotations in a \java program. Therefore $\textbf{$\#$ LOC(M)$_{P}$} = \textbf{N} + \textbf{C} + 1$.
The number of permission contracts (\code{N}) would be equal to the number of methods \textbf{M} if analysis generates permission contracts for all the methods. 

\item[\textbf{$\#$ Anns(M)$_{P}$}] in Table \ref{table:annotation} calculates the number of individual annotations for \textbf{M} methods in the \pulse translated version of the each program. The number of annotations in this case depends on the presence (absence) of annotations on the receiver object (\code{this}) and the number of parameters (aliases of the referenced objects) accessed in a method.

For each non-constructor method accessing a field of a referenced object \code{(this)}, it includes two annotations to specify pre- and post-permission on the object.

\begin{table}[htb]
\begin{center}
\caption{Effectiveness and efficiency analysis of the permission inference technique.}
\label{table:annotation}
\begin{threeparttable}
\small
\begin{tabular}{p{40pt}cccp{25pt}}
\hline\noalign{\smallskip}
\multicolumn{1}{c}{ } & \multicolumn{3}{c}{Annotation Overhead}  & Performance \\
\cmidrule(lr){2-4} \cmidrule(lr){5-5}
\textbf{Program}&\textbf{$\#$ LOC(M)$_{P}$}\tnote{$\sigma$}&\textbf{$\#$ Anns(M)$_{P}$}\tnote{$\sigma$}&\textbf{$\#$ Anns(M)$_{F}$}&\textbf{Time(M)} (sec)\\
\noalign{\smallskip}\hline\noalign{\smallskip}
crystal&2,188&5,234\tnote{$\triangleleft$}&6,691\tnote{$\triangleleft$}&1441.056\\
\midrule
pulse&513&1,764&4,850&41.84\\
\midrule
montecarlo&204&948&1,360&17.95\\
euler&52&197&1,073&5.42\\
search&60&204&691&1.41\\
moldyn&52&205&901&1.40\\
lufact&50&325&437&1.13\\
crypt&46&163&385&0.98\\
series&43&143&207&0.56\\
sor&42&153&267&0.61\\
sparsematmult&36&156&316&0.51\\
\midrule
blacksholes&56&250&694&1.92\\
gaknapsack&38&82&250&1.03\\
health&25&74&334&0.30\\
\midrule
webserver&11&25&11&0.24\\
fft&16&56&44&0.43\\
quicksort&13&33&17&0.06\\
shellsort&10&32&44&0.06\\
integral&7&17&17&0.08\\
fibonacci&8&18&9&0.03\\
\midrule
ArrayCollection&15&78&49&0.16\\
\midrule
\textbf{Total}&\textbf{3,485}&\textbf{10,157}&\textbf{18,647}&\textbf{1,517.47}\\
\noalign{\smallskip}\hline
\end{tabular}
\begin{tablenotes}
\item[$\sigma$]{\scriptsize{The parameter notations are not included in this count.}}
\item[$\triangleleft$]{\scriptsize{It includes 2,324 annotations for the 581 overloaded methods with at least 4 annotations per method.}}
\end{tablenotes}
\end{threeparttable}
\end{center}
\end{table}
For each non-constructor method, \textbf{Anns(M)$_{P}$} counts two (as part of \texttt{requires} and \texttt{ensures} clause) to add a typestate `\code{alive'} as a part of pre- and post-permission for each class field and (or) parameter at the method level.

For each non-constructor method with parameters, \textbf{Anns(M)$_{P}$} adds two (pre- and post-permission) for each parameter.

For each \emph{non-constructor} method that returns a global object or alias of a global reference, \textbf{Anns(M)$_{P}$} adds two annotations as a part of \texttt{ensures} clause to specify the permission generated by a method on the returned object ($R$). Finally, \textbf{Anns(M)$_{P}$} counts one for the annotation \code{ENDOFCLASS} to mark the end of each class in a program (as required by \pulse).

For each constructor method, there is only one permission annotation for receiver object (\code{this}) and one typestate annotation. It can have only `\texttt{ensures}' clause as part of the permission contract in \pulse. There are no annotations for the parameters as \pulse does not support overloaded constructors with parameters.

Let $M$ be the set of methods, \textbf{M$_C$} be the number of constructors, \textbf{M$_{NC}^F$} be the number of non-constructor methods that access some global (class) field directly or indirectly in a method, \textbf{M$_{NC}^R$} be the number of non-constructor methods that returns a referenced object
(i.e., $\textbf{M} = \textbf{M}_C + \textbf{M}_{NC}^F$) + \textbf{M$_{NC}^R$}. \textbf{Anns}$_{P}$ is defined as:
 \begin{equation}\label{eq:eq3}
   \textbf{Anns}_{P} = (1 + 1) * \textbf{M}_C  + (2 + 2) * \textbf{M}_{NC}^F + \sum_{m \in M}(2 * P(m)) + 2 * \textbf{M}_{NC}^R + \textbf{C}
 \end{equation}
where $P(m)$ denotes the number of parameters (aliases of some global object) accessed in method $m$.
    
\item[\textbf{$\#$ Anns(M)$_{F}$}] calculates the number of permission-based annotations (pre- and post-permissions)
generated on individual fields (\code{F}) of the current object (\code{this}) for \code{M} methods in a program. The numberof annotations also includes two annotations (generated as post-permission) on the returned object ($R$) by a non-constructor method. Therefore, the number of permission-based annotations generated at field level for \textbf{M} methods is $\textbf{Anns(M)$_{F}$} = \sum_{m \in M}(2 * {F} (m))$ + $2 * \textbf{M}_{NC}^R$, where $F(m)$ is the number of individual fields and its aliases accessed in method $m$ and $\textbf{M}_{NC}^R$ be the number of methods that returns a referenced object.

In total, \gap generated 18,647 permission annotations at the field level, for 3,111 methods in 350 classes for 20 benchmark programs and 3,485 permission annotated lines were generated for the \pulse translated versions of the same programs with 10,157 minimum annotations in less than 1,517 seconds (about 26 minutes). 

For example, for the \pulse translated version of the example program given in Listing \ref{lst:sa-specprog} with three classes and 12 methods, \gap generated 15 annotated lines with a total of 78 annotations at object level and 49 annotations when permissions were generated on individual fields of the referenced object. Likewise, for \crystalsaf, \gap generated 2,188 annotated lines with a minimum of 5,234 annotations, for 1,394 methods for its \pulse translated version. For \crystalsaf, we omit the parameter annotations from the annotated version to avoid the state space explosion problem due to the model checking of the specifications in \pulse but that does not invalidate the effectiveness of our technique in generating specifications without annotation overhead.
\end{description}

The analysis shows the effectiveness of the inference technique in automatically generating the specifications which otherwise needs to be manually identified and added in the program and that pose a significant annotation overhead for the programmers.

\subsection{Performance analysis}
We compute efficiency of the permission inference algorithm by measuring the average execution time of the analysis in seconds, averaged over 10 independent runs on \jvm. Table \ref{table:annotation}, column Time(M), shows the execution time (in seconds) of the analysis to parse the source code and infer permissions for \code{M} methods in a program.
The result shows that it takes about 18 seconds to generate specifications for montecarlo program with 196 methods, and a fraction of a seconds for programs in the \aem and \plaid benchmarks being smallest in size. For \crystalsaf, the biggest case study we evaluated with 17K plus physical source lines of code, it took 1441.056 seconds (about 25 minutes) to identify the permission-based dependencies from the source code and generate 2,188 annotated lines of code with a minimum of 6,691 annotations for 1,394 methods when permission are generated at the field level.

If created manually, the above annotations may take time that could be multiples of months as in the case of \pulse~\cite{CatanoASA14} itself where authors reported that they took six months to understand (identify) dependencies and manually annotate a \java application with 55 classes and 376 methods with 546 permission contracts to verify program behavior. No doubt, the complexity and the design of the input program does have an impact on the annotations generation time. However, we expect that the execution time of \gap can be a multiple of minutes (or hours in the worse-case for really large applications), but not in months, thus showing the effectiveness of our technique in generating permission-based annotations while saving time.

\section{Implications}\label{implications}
We observe, that our proposed framework, \gap, by automatically inferring access permission contracts from a source program not only helps programmers to work at a higher level of abstraction letting them focus on the functional and behavioural correctness of the program, but can also be used to automatically identify some of the syntactical errors in the program, such as \emph{null-pointer} references, without actually compiling the program. Further, having contract-based specifications in an explicit way can help programmers to identify dependencies at the method level and compute the potential for concurrency in a sequential program.

We now revisit the motivating example, given in Listing \ref{lst:sa-specprog}, to demonstrate the efficacy and expressiveness of permission-based specifications in identifying some of the program properties such as null pointer exceptions and enabling implicit concurrency present in a sequential program.

\begin{description}[leftmargin=0cm]
\setlength{\itemsep}{4pt}
\item \textbf{Null Pointers Analysis.}
With respect to the permission semantics, the null-pointer exceptions can arise in two ways in a program: (a) program error: reference to an object is a \code{null} reference itself and, (b) permission inference error: no method generates the permission required of a method, say \code{unique}, on the referenced object as its post-permission. For example, in Listing \ref{lst:sa-specprog}, all the methods accessing the shared object, say \code{array1}, would cause a \code{null-pointer} exception and would remain unreachable, if the client method, (in this case the constructor method), does not generate \code{unique} permission on its referenced object before using it. Similarly, the post-permissions (\code{none(array1)} and \code{none(array2)} of method \code{tidyupColl()} indicate that object \code{array1} and \code{array2} should be instantiated again, once the method \code{tidyupColl()} has been executed, for these objects to be used by other methods without generating the \code{null-pointer} exception. The proposed framework can identify such subtle errors in the program based on the inferred specifications.

\item \textbf{Method-level Dependencies Analysis.}
Further, access permission contracts can be used to compute the dependencies between methods and to automatically impose ordering constraints.

For example, a close examination of the \code{requires} clause for method \code{tidyupColl()} in Listing \ref{lst:sa-specprog} (Line 26), shows that the method can only be called if objects \code{array1} and \code{array2} has \code{unique} permission as pre-permission. This means it can either be called immediately after the constructor methods \code{ArrayCollection(array1)} and \code{ObjectClass(array2)}, as both generate its required permissions, or once all the methods accessing the \code{array1} and \code{array2} object have completed their execution, and \code{unique} permission on the referenced objects have been resumed, to be used by other methods.

\item \textbf{Implicit Concurrency Analysis.}
Furthermore, having access permission contracts defined, in an explicit way, at the method level, can be used to compute the number of concurrent methods and the methods that should always run sequentially. 
as elaborated in Section \ref{conc-analysis} and that is also evident from the study of the existing permission-based parallelization approaches \citep{aldrich2012plaid,Aeminium2014}. 

For example, the method call concurrency graph (Figure \ref{fig:dependencygraph-example3}) given in Listing \ref{lst:sa-specprog}, reveals the effectiveness of \gap framework in generating access permissions at the field level, that can help achieve concurrency at a more granular level, than generating permissions on the whole object. It shows that 11 out of 15 method calls in this program can be executed in parallel, based on the inferred (permission) contracts. Recall the \pulse concurrency analysis for the same program in Figure \ref{fig:example-summary} that reports 4 out of 11 methods to be concurrent. This is because in \pulse, the concurrency analysis of the program is based on the access permission contracts defined at the object level.


\item \textbf{Document Generation.}
Moreover, the \gap framework automatically generates a user-friendly analysis (.pdf visualization) report
that provides developers a quick, abstract and explicit view of the implicit relations (dependencies) that exists at the method level and their concurrent behavior without looking at the source code. The generated report can be used by both novice and expert programmers (verifiers) with equal ease to automatically analyze program behaviour of the program in terms of its null-pointer and code reachability analysis, without performing any code inspection.
\end{description}

We believe, that the automatic inference of access permission contracts and its concurrency analysis by the \gap framework, along with its integration with the model checking verification tool, \pulse, opens a new window for the programmers (verifiers) to automatically verify the intended behavior of large and complex applications and to parallelize the execution of sequential programs, without any extra effort (annotation overhead) and in a safe way with less time.
\section{Threats to Validity}\label{threats}

We have identified some limitations of the \tool{} framework and threats to the validity of the underlying permission inference and checking approach itself.
\begin{description}[leftmargin=0cm]
\setlength{\itemsep}{5pt}
\item[\textbf{Construct Validity.}]
The permission extractor in \gap, at the moment, does not support all the language constructs in \javasl such as method overriding (polymorphism), generics and lambda expressions. This is because the technique is based on statically analyzing the source code. However, we believe the approach can be extended to identify the overridden method called, as a result of a method call, without actually performing the dynamic analysis of the program, as we perform flow-sensitive analysis of the source code to determine the actual instance or type of object assigned to a reference of base type in an assignment expression. Moreover, our evaluation on small to large sized benchmark programs validate that through \gap, we have provided a succinct methodology to infer access permissions for realistic \java programs, implementing rich constructs of \javasl.

\item[\textbf{Internal Validity.}]
In certain situations, the permission inference framework does not produce the optimal (precise) solutions and generates safe (restrictive) permissions i.e., \code{unique} or \code{full} instead of \code{immutable} or \code{pure}.

This happens when a) an object is accessed in a complex infix expression with nested pre-fix or post-fix expressions or b) an object is accessed in a library method call expression. The first case is because the analysis is based on the type of expressions such as \code{Field\_Access} and \code{Assignment} encountered in the \code{AST} of program, therefore, the permission extractor does not individually parse complex \code{infix-postfix} expressions as a part of the implementation. The second case is due to the unavailability of method definitions (bodies) for the library methods.

We believe this is an engineering problem and generating safe (restrictive) permissions will not affect the integrity of the program itself when actually used for program verification or parallelization purpose. 

\item[\textbf{External Validity.}]

It is worth noting here that the \tool{} framework, at the moment, does not automatically check the correctness of the inferred specifications for the overloaded constructors (methods) and consequently, it does not include them as a part of its concurrency analysis. We observe, the same is true for the parameter annotations, generated as a part of the method contracts, in the \pulse translated version of the input program.
\begin{itemize}[leftmargin=0cm]
\setlength{\itemsep}{4pt}
\item[-] This is because the \pulse tool does not support overloaded methods and parameter annotations as a part of its correctness and concurrency analysis. Consequently, the \gap framework excludes the overloaded constructors (methods) from the \pulse translated version of the program and analyzes them manually. However, the actual count of the concurrent methods may be misleading (underestimated) for programs having overloaded methods.
In the case of parameters, the analysis (explained previously in Section \ref{meta-analysis}, internally handles this problem by mapping all the formal parameters with their actual references and generating permissions on the receiver object that are then verified through the \pulse tool.

\item[-] Further, the concurrency analysis of the input specifications in the \pulse tool is performed at the class level, which makes it difficult for the \gap framework to check the potential for concurrency at the project level. However, it does not invalidate the effectiveness of the permission inference approach, in generating the correct specifications without annotation overhead as the problem is due to the inherent limitation of the \pulse tool itself.
\end{itemize}

We believe all the limitations discussed above are due to the inherent limitation of the \pulse tool and are engineering problems that can be solved, by extending the \pulse analysis with overloaded methods, and parameter annotations. However, in both cases, we manually check the inferred specifications as explained in Section \ref{quality}.

\item[\textbf{Termination Analysis.}]
The termination of the permission inference analysis in the \gap tool is based on the following characteristics.
\begin{itemize}[leftmargin=0cm]
\setlength{\itemsep}{4pt}
\item[-] The analysis is based on the \code{AST} of the source code, a parse tree having a finite number of nodes (sub-expressions) for an expression statement. Therefore, it takes a finite number of steps to parse the generate tree with a finite number of expressions.
Therefore, the proposed analysis always converges to an expression type, designated as the base case, or otherwise terminates successfully.

\item[-] The presence of indirect or chained recursion, as explained previously in Section \ref{meta-analysis}, can cause indefinite (infinite) loops and consequently, the memory overflows. This is because we save the meta-data (at-least its signature) of each method, and its current state, before parsing its body and switching the control to parse sub-method calls in it. This step helps to identify the second level (indirect) recursive method call for the same method during parsing, and to ensure that analysis terminates successfully. 

\item[-]  Another threat for the successful termination of the analysis can be the expression statements designed as \code<self-address-flow> statements (when a reference directly or indirectly starts pointing to itself creating a loop for analysis), the technique identifies such expressions following the pre-defined syntactic rules, (Appendix \ref{rules}) and terminates the analysis successfully, without creating any loop or cycles during the permission extraction process.
\end{itemize}

\item[\textbf{Soundness Analysis.}]
In the context of permissions, the analysis is sound if it generates the required or safe permissions for each memory location that a method needs to read or write on during its execution.

The \gap tool ensures to generate the required or sufficient permissions, we call it safe permissions. For this purpose, the analysis is supported by the mathematically specified syntactic rules (Appendix \ref{rules}), that help to precisely model the object's accesses in a permission-based graph model and generate permissions on the referenced objects. Further, the underlying approach ensures to minimize the number of false positives, which means that the \gap framework always generates permissions when and where required.

Moreover, evaluation of the \gap framework on realistic \java applications and its proof-of-concept by integrating (extending) the state-of-the-art model checking approach, \pulse, demonstrates the soundness (at least precision) of the underlying approach in generating the correct and required permissions.
\end{description}
\section{Related Work}\label{relatedwork}
For the sake of brevity, we will discuss here the most relevant permission-based concurrent programming paradigms such as \plaid \cite{aldrich2012plaid,aldrich2011plaid} and \aem \cite{StorkMA09,Aeminium2014}, the program verification approaches \cite{bierhoff2007modular,Bierhoff-plural-tool,Bierhoff09polymorphicfractional,Chalice2009,Amighi2012TheBasecamp,Amighi2014,Juhasz2014Viper,Huisman2015AReasoning,Walter2016AutomaticInterpretation} and the inference of access permissions \cite{heule2013abstract,Ferrara2012,Dohrau2018PermissionPrograms} presented in the literature.

\plural (Permissions Let Us Reason about Aliases) \cite{bierhoff2007modular} is a formal specification language and a type system. The language was initially designed to verify correctness of API protocols such as JDBC, iterators in \java, and their compliance in sequential and concurrent programs based on permission-based typestate information where access permissions are used to keep track of multiple references to a particular object and typestate \cite{strom1986typestate} defines the valid sequences of operations that can be performed upon an instance of a given type. 
The technique was implemented as a \java Eclipse plug-in \cite{Bierhoff-plural-tool}. It performs intra-procedural static analysis, called DFA (Diagram Flow Analysis) of the source code to identify and track access permissions and typestate information associated with each program variable (method parameters, the method receiver object
and local variables) at every program point and issues warning for protocol
violations (if any) at runtime. 

In an extended work to \plural, Bierhoff et al. \cite{Bierhoff09polymorphicfractional} proposed a technique to infer access permission flow in the system based on the input specifications.
The objective was to avoid permission tracking overhead associated with splitting and joining the fractional permission values during program execution. No doubt, \plural helps programmers to statically
follow API protocols and to ensure their compliance, without actually executing the program. However, to perform program verification, programmers are required to
explicitly specify permission-based typestate annotations as a part of methods' contract which creates annotation overhead for programmers.

Chalice \cite{leino2009basis} is a permission-based specification language and a program verifier \cite{Chalice2009} that uses concept of Boyland's fractional permissions to reason about the non-interference of shared state in a Chalice program. It defines permission percentages as `Full',`Some' and `No', in the range (0, 100], to specify the concurrent reading, (un)sharing of objects among multiple threads, for each heap location in the program. However, in Chalice, programmer manually annotates program with accessibility predicates, comprising permission-based specifications as monitor (class) invariants, to verify its correctness.
In a chalice program, the annotation acc(o.f) represents
`Full' permission (100\%) that shows that a thread
has exclusive access to the field (f). A non-zero or `Some' permission
shows a read-only access to a particular location (o.f) that is represented
by rd(o.f).
The permission contracts are transferred
from a monitor to a thread that acquires this monitor.
The approach ensures that the sum of permissions
from all threads remain less than or equal to 100\%.
Chalice does not support automatic inference
of access permissions but it uses autoMagic, a command-line option, to infer 'access' and `pure' assertions as read and write accesses for all the non-heap locations in a program.

\javasyp (Symbolic Permissions for efficient static program verification) \cite{bierhoff2011automated} is a permission-based automated program verification tool that verifies functional correctness of sequential programs such as \java array list with its iterator.
The underlying approach combines symbolic permissions (unique and immutable) with \jml contracts enforce and control aliasing in the program. The technique performs static analysis of the source code 
and generate verification conditions (VCs) based on permission contracts. The inferred conditions are then used to verify program behavior. In this approach, permission tracking
is straightforward as tracking symbolic values is much easier than fractions of values. However, it requires programmers to explicitly specify permission-based specifications (class invariants and other necessary conditions) in the program that creates annotation overhead for the programmers.

As discussed previously, \pulse~\cite{SiminiceanuAC12} is an automatic formal verification tool that performs static analysis of a \plural annotated \java program to verify correctness of the input specifications.
Cata{\~n}o et al. \cite{CatanoASA14} used \pulse to verify correctness of a multi-threaded \java application MTTS (Multi-threaded task server) by manually adding \plural specifications in the program. MTTS is a multi-threaded task distribution server, developed by Novabase that parallelizes computational tasks over multiple threads. 
In \pulse, programmers specify behavioral properties (design intents) of the program as permission-based locking information and typestate invariants to enforce the properties that should hold true during program execution.
The authors reported that it took six months to understand the implicit dependencies in the program and to manually annotate a \java program with 546 annotated lines of \plural specifications which poses annotation overhead for the programmers. 


VeriFast \citep{jacobs2011verifast} is a sound, modular and automatic program verification tool for single and multi-threaded C and \java programs. However, to enable verification, programmers interactively specify the access permission contracts using fractional permissions in the range $(0,1]$, by defining Lemma functions in the program written in classic Separation Logic \cite{reynolds2002separation}. The permission value 1 specifies the exclusive (write) access for manipulating a heap location and any smaller value less than 1 is used to specify the concurrent read access for an object between multiple references. Lemma functions are like ordinary C functions, except that lemma functions and calls for lemma functions are written inside annotations. The logic-based specifications are then tracked in the system to ensure that, in a permission contract, pre-conditions imply post-conditions and verify program behavior based on input specifications.

\renewcommand*{\thefootnote}{\arabic{footnote}}
VerCors\footnote{\url{https://fmttools.ewi.utwente.nl/redmine/projects/vercors-verifier/wiki/Puptol/}}
\citep{Amighi2012TheBasecamp,Amighi2014},
is an automated verification tool that verifies correctness of concurrent data structures based on fractional permissions.
The verification approach in VerCors 
extends Silicon \citep{Juhasz2014Viper} as a back-end verification tool that natively supports an expressive permission model. VerCors supports program written in \java and OpenCL\footnote{Khronos OpenCL Working Group, The OpenCL specification, \url{http://www.khronos.org/opencl/}}. The approach exploits verification capabilities of \jml \citep{leavenspreliminary06} to support multiple synchronization primitives and to reason about the functional correctness of an input program. However, to perform verification, it requires programmers to explicitly define permission-based invariants as \jml comments in the program.

In VerCors, permissions are specified using propositional formulae of the form \code{Perm(e.f, $\pi$)} at the method level, where $\pi$ shows the fractional permission, in the range (0, 1], associated with the field $f$ of an object $e$. The permission are then transferred between threads at synchronization points and tracked with the execution of the program. However, instead of simply defining the amount (in fractions) of the permission transferred, \citet{Huisman2015AReasoning} extended the fractional permission model in VerCors, by annotating program with
symbolic expressions that define the kind of \textit{transfer} applied to a permission and the owner of the transferred permissions thereby, using a higher-level of abstractions compared to Veri-Fast and Chalice.

Viper \cite{Juhasz2014Viper,Muller2017Viper:Reasoning} is a verification infrastructure that encodes permission reasoning in an object-based intermediate language. The infrastructure
includes two back-end verifiers and four front-end tools for Chalice, \java, Scala, and
OpenCL that was developed as a part of VerCors project \citep{Blom2014ThePrograms}. It targets a sequential,
object-based intermediate language and verifies heap structures in a program based on the permissions contracts. Like Chalice, in a Viper program, access rights are specified using accessibility predicates \citep{Cousot1977} that define permission-based concrete and recursive specifications as pre- and post-conditions and the loop invariants for heap structures. A method can access a particular heap location \code{e.f} if appropriate permissions are held by that location. The annotation \code{acc(e.f)} specifies \code{full} permission for a memory location \code{e.f} that needs to be updated while a permission greater than zero specifies a read access for a particular memory location. The annotated program is then encoded into an intermediate language by Viper's front-end tools to be verified by the back-ends tools. 
In a Viper program, access permissions are held by a method execution or in the loop body that are then transferred between method execution and the loop invariants to verify program behavior and program verification is based on the input specifications defined at the method level rather than program implementation.

\citeauthor{Ferrara2012} in \cite{Ferrara2012} developed a technique to infer access permission rights to verify correctness of concurrent programs for a small object-based language. The technique performs static analysis of the source program to infer fractional and counting permissions based on abstract interpretations \citep{Cousot1977}, a theory for defining and soundly approximating the semantics of a program. 
However, to infer permissions, programmer manually annotate program with the access notations as pre- and post- conditions at method level. 

The approach firstly computes symbolic values (approximations) for each heap location using the pre- and post-conditions at method. It then infers constraints over these symbolic values to reflect permission-based intermediate representations for each heap location. Finally, it generates specifications in the form of fractional (value between 0 and 1) and counting (value between 0 and $Integer:Max\_Value$) permissions using the inferred constraints. The inference technique is implemented in \code{Sample} (Static Analyzer of Multiple Programming LanguagEs)\footnote{\url{http://www.pm.inf.ethz.ch/research/sample.html}}
that supports programs written in Scala \citep{scalaodersky2004}.  However, it is generally acknowledged that specifications based on fractions (concrete) values are tedious to write and harder to track and adapt for programmers ~\cite{heule2013abstract}.

Heule et al.~\cite{heule2013abstract} proposed a technique that automatically converts fractional permissions for shared-memory concurrent programs into abstract \code{read} and \code{write} permission. The objective was to avoid complex reasoning overhead to specify concrete (fractional) values for concurrent constructs in a program.
This permission system generates two kinds of access notations i.e., \code{full} and \code{read}, where \code{full} represents exclusive access rights and \code{read} represents the read-only access to a shared variable. Like other approached, the proposed approach takes an annotate program with accessibility predicates (access \& read) as a part of method specifications to infer permission.

Walter \cite{Walter2016AutomaticInterpretation} proposed a static analysis to infer quantified permissions and loop invariants for arrays and recursive data-structures in Viper. Like Viper, the program is first annotated with accessibility predicates such as \code{acc (loc(array, index).val)} to specify the access rights (\code{acc}) for an array index (memory location) and the loop invariants. The technique identifies the array indices accessed in a program, through a read or write operations, and generates logical formulas for all these locations. The locations are then analyzed by the Viper infrastructure that generates quantified permissions as weakest pre-condition for the referenced objects. The generated permissions are in the form of fractional permissions where 1 represents \textsf{full} permission and any positive fraction of permission greater than 0, represents the read access. The technique extends input program with ghost parameters (functions) to infer read permissions for the specified locations. However, the analysis does not keep account of the aliasing information as it does not allow the expressions on right hand side of an assignment statement to be heap dependent therefore, at the moment, the approach does not handle complex data-structures such as arrays of arrays (2D arrays). Moreover, it is reported that the approach generates lengthy and complex specifications that are sometime difficult to read and understand by a human.

Following the initial idea of Walter's permission inference mechanism for concurrent programs, Dohrauet et al. in \cite{Dohrau2018PermissionPrograms} proposed a static analysis to infer permission-based contracts for array manipulating concurrent programs. The analysis supports the programming languages that supports ownership transfer. The idea is to associate a separate permission with each array element based on the classic Separation Logic \citep{reynolds2002separation}. The technique employs fractional permissions to ensure the exclusive (full) write access on a particular location while allowing the concurrent (read) access using any positive fraction of permission. This time the inferred specifications are human readable but it still does not support arrays of arrays. The technique is written in Scala and requires an input program written in the Viper language with user-provided annotations.


\plaid~\cite{aldrich2011plaid} is permission-based programming language that verifies and parallelise execution of typestate-based sequential programs based on access permissions. It defines permission-based typestate contract as a part of language. Every type in \plaid is explicitly represented as a tuple having a type structure and associated permission to show aliasing and mutability of the corresponding object type. It uses \textsf{unique} and \textsf{immutable} to define permission for individual objects, and \textsf{share} permission is used to define the access rights for the related (shared) objects. \plaid's run system time then dynamically track the access permission flow in the program to verify its behavior. 

Having a language and run time support for access permissions in \plaid, Stork et al.~\cite{Aeminium2014}\
developed a new programming paradigm \aem to develop by-default concurrent applications based on access permissions. 
Like \plaid, it uses three kinds of access permissions: \textsf{unique}, \textsf{immutable}, and \textsf{share}.
\aem leverages access permission flow through the system to compute data dependencies at the task level and parallelise execution of the program as permitted by the computed dependencies. \aem is able to achieve significant performance improvements, but at the cost of increasing programmers' burden to explicitly specify statefull effects in the form of access permission rights at method level.

Although, all the existing permission-based approaches have been quite promising in achieving their desired goals. However, the approaches are either based on a) formal type theories or programming models or b) use some intermediate representations and (or) c) generate permissions in low level (concrete) values based on the provided annotations at the code level. 

In our work, we automatically infer symbolic permissions from the programs written in mainstream programming language (\java) without using any intermediate representations and method-level specifications. We observe that relieving programmers from the complex specification (annotation) overhead, the aim of our research, is applicable to both the program verification and parallelization approaches
We believe that inference of access permissions can be the next step to exploit the verification and parallelization power of existing approaches and to enhance their applicability for the mainstream programmers and verifiers. Moreover, the inference of access permissions in the form of higher level of abstractions can free programmers from the low-level analysis and tracking overhead associated with handling concrete values in the program.

\section{Conclusion}\label{conclusion}
To exploit parallelism offered by multi-core processors, mainstream programming languages such as Java typically make use of explicit concurrent programming constructs such as threads and locks. However, such constructs give rise to significant code complexity and errors. 
Therefore, access permissions-based dependencies have been proposed as an alternative approach to achieve implicit concurrency without using explicit concurrency constructs. However, these specifications need to be manually added in the source program imposing significant additional work for programmers which is time consuming and error-prone.

Therefore, in this paper, we propose, \gap, a framework to automatically inferring permission-based implicit dependencies in the form of symbolic permissions from sequential \java programs by performing static analysis of the source code. Moreover, we integrated and extended \pulse, a permission-based model checking tool to perform empirical evaluation of the inferred specifications and to reason about their concurrent behavior.
Experiments were performed on 20 programs with 3,111 methods from four widely-used benchmark datasets, \jomp, \aem, \plaid, \crystalsaf and \pulse itself to evaluate the correctness and effectiveness of the inferred annotations, as well as the efficiency of our inference algorithm. The concurrency analysis of the inferred specifications using \pulse and the \gap confirms that on average, 61\% methods can be parallelised with each other based on the inferred contracts in all the benchmark programs. Our algorithm also infers permission-based annotations efficiently, averaging 2 seconds for annotating a single method. Moreover, all the experiments demonstrate the feasibility and benefits of our framework to the related permission-based verification and parallelization approaches in the literature to achieve their intended benefits without imposing additional work (annotation overhead) for programmers.

We observe that the inference of permission contracts can further be used to automatically infer the synchronization primitives (locking and ordering constraints) from the source code
of sequential programs and to parallelize program execution for the mainstream programming languages such as Java, without using any new programming language and run time system to support access permissions. Further, the inference of permission-based synchronization constructs (such as acquire
and release locks with permission invariants) can be used to verify the behavior of already concurrent
programs without imposing extra work
on the programmer side. 

To this end, we have envisaged a number of future directions relevant to the proposed framework. We plan to (a) expand the inter-procedural static analysis to incorporate more \java language constructs such as polymorphism, generics, lambda expressions and others; (b) infer access permissions at a more granular level, such as individual permissions for members of collections or arrays; (c) develop an online system to encourage the wider adoption of the proposed technique; (d) automatically infer the permission-based locking and ordering constraints to develop by-default concurrent application in \java; (e) extend \pulse analysis to overcome its existing limitations and provide a comprehensive support for the \javasl; and (f) integrate \plural, \pulse and \gap to develop an integrated framework for permission-based automatic program verification for \java.

\section*{Acknowledgment}
This research was conducted under the Endeavour Leadership Award for Ph.D. and funded by the Department of Education and Training, Government of Australia and Faculty of Information Technology at Monash University, Australia. 

\bibliographystyle{elsarticle-harv}
\bibliography{library.bib}

\begin{thebibliography}{47}
\expandafter\ifx\csname natexlab\endcsname\relax\def\natexlab#1{#1}\fi
\providecommand{\url}[1]{\texttt{#1}}
\providecommand{\href}[2]{#2}
\providecommand{\path}[1]{#1}
\providecommand{\DOIprefix}{doi:}
\providecommand{\ArXivprefix}{arXiv:}
\providecommand{\URLprefix}{URL: }
\providecommand{\Pubmedprefix}{pmid:}
\providecommand{\doi}[1]{\href{http://dx.doi.org/#1}{\path{#1}}}
\providecommand{\Pubmed}[1]{\href{pmid:#1}{\path{#1}}}
\providecommand{\bibinfo}[2]{#2}
\ifx\xfnm\relax \def\xfnm[#1]{\unskip,\space#1}\fi
\bibitem[{Aldrich et~al.(2012)Aldrich, Beckman, Bocchino, Naden, Saini, Stork
  and Sunshine}]{aldrich2012plaid}
\bibinfo{author}{Aldrich, J.}, \bibinfo{author}{Beckman, N.E.},
  \bibinfo{author}{Bocchino, R.}, \bibinfo{author}{Naden, K.},
  \bibinfo{author}{Saini, D.}, \bibinfo{author}{Stork, S.},
  \bibinfo{author}{Sunshine, J.}, \bibinfo{year}{2012}.
\newblock \bibinfo{title}{{The Plaid language: Typed core specification}}.
\newblock \bibinfo{type}{Technical Report}. DTIC Document.
\bibitem[{Aldrich et~al.(2011)Aldrich, Bocchino, Garcia, Hahnenberg, Mohr,
  Naden, Saini, Stork, Sunshine, Tanter and {others}}]{aldrich2011plaid}
\bibinfo{author}{Aldrich, J.}, \bibinfo{author}{Bocchino, R.},
  \bibinfo{author}{Garcia, R.}, \bibinfo{author}{Hahnenberg, M.},
  \bibinfo{author}{Mohr, M.}, \bibinfo{author}{Naden, K.},
  \bibinfo{author}{Saini, D.}, \bibinfo{author}{Stork, S.},
  \bibinfo{author}{Sunshine, J.}, \bibinfo{author}{Tanter, E.},
  \bibinfo{author}{{others}}, \bibinfo{year}{2011}.
\newblock \bibinfo{title}{{Plaid: a permission-based programming language}},
  in: \bibinfo{booktitle}{Proceedings of the ACM international conference
  companion on Object oriented programming systems languages and applications
  companion}, \bibinfo{organization}{ACM}. pp. \bibinfo{pages}{183--184}.
\bibitem[{Amighi et~al.(2014)Amighi, Blom, Darabi, Huisman, Mostowski and
  Zaharieva-Stojanovski}]{Amighi2014}
\bibinfo{author}{Amighi, A.}, \bibinfo{author}{Blom, S.},
  \bibinfo{author}{Darabi, S.}, \bibinfo{author}{Huisman, M.},
  \bibinfo{author}{Mostowski, W.}, \bibinfo{author}{Zaharieva-Stojanovski, M.},
  \bibinfo{year}{2014}.
\newblock \bibinfo{title}{{Verification of Concurrent Systems with VerCors}},
  in: \bibinfo{booktitle}{Formal Methods for Executable Software Models: 14th
  International School on Formal Methods for the Design of Computer,
  Communication, and Software Systems}. \bibinfo{publisher}{Springer
  International Publishing}. SFM'14, pp. \bibinfo{pages}{172--216}.
\bibitem[{Amighi et~al.(2012)Amighi, Blom, Huisman and
  Zaharieva-Stojanovski}]{Amighi2012TheBasecamp}
\bibinfo{author}{Amighi, A.}, \bibinfo{author}{Blom, S.},
  \bibinfo{author}{Huisman, M.}, \bibinfo{author}{Zaharieva-Stojanovski, M.},
  \bibinfo{year}{2012}.
\newblock \bibinfo{title}{{The {VerCors} Project: Setting Up Basecamp}}, in:
  \bibinfo{booktitle}{Programming Languages meets Program Verification (PLPV
  2012)}.
\bibitem[{Beckman(2009)}]{beckman2009modular}
\bibinfo{author}{Beckman, N.E.}, \bibinfo{year}{2009}.
\newblock \bibinfo{title}{{Modular typestate checking in concurrent Java
  programs}}, in: \bibinfo{booktitle}{Proceedings of the 24th ACM SIGPLAN
  conference companion on Object oriented programming systems languages and
  applications}, \bibinfo{organization}{ACM}. pp. \bibinfo{pages}{737--738}.
\bibitem[{Beckman et~al.(2008)Beckman, Bierhoff and Aldrich}]{Beckman:2008}
\bibinfo{author}{Beckman, N.E.}, \bibinfo{author}{Bierhoff, K.},
  \bibinfo{author}{Aldrich, J.}, \bibinfo{year}{2008}.
\newblock \bibinfo{title}{{Verifying Correct Usage of Atomic Blocks and
  Typestate}}, in: \bibinfo{booktitle}{Proceedings of the 23rd ACM SIGPLAN
  Conference on Object-oriented Programming Systems Languages and
  Applications}, \bibinfo{publisher}{ACM}. pp. \bibinfo{pages}{227--244}.
\bibitem[{Bierhoff(2011)}]{bierhoff2011automated}
\bibinfo{author}{Bierhoff, K.}, \bibinfo{year}{2011}.
\newblock \bibinfo{title}{{Automated program verification made SYMPLAR:
  symbolic permissions for lightweight automated reasoning}}, in:
  \bibinfo{booktitle}{Proceedings of the 10th SIGPLAN symposium on New ideas,
  new paradigms, and reflections on programming and software},
  \bibinfo{organization}{ACM}. pp. \bibinfo{pages}{19--32}.
\bibitem[{Bierhoff and Aldrich(2007)}]{bierhoff2007modular}
\bibinfo{author}{Bierhoff, K.}, \bibinfo{author}{Aldrich, J.},
  \bibinfo{year}{2007}.
\newblock \bibinfo{title}{Modular typestate checking of aliased objects}, in:
  \bibinfo{booktitle}{Proceedings of the 22Nd Annual ACM SIGPLAN Conference on
  Object-oriented Programming Systems and Applications},
  \bibinfo{publisher}{ACM}. pp. \bibinfo{pages}{301--320}.
\bibitem[{Bierhoff and Aldrich(2008)}]{Bierhoff-plural-tool}
\bibinfo{author}{Bierhoff, K.}, \bibinfo{author}{Aldrich, J.},
  \bibinfo{year}{2008}.
\newblock \bibinfo{title}{{PLURAL: Checking Protocol Compliance Under
  Aliasing}}, in: \bibinfo{booktitle}{Companion of the 30th International
  Conference on Software Engineering}, \bibinfo{publisher}{ACM}. pp.
  \bibinfo{pages}{971--972}.
\bibitem[{Bierhoff et~al.(2009a)Bierhoff, Beckman and
  Aldrich}]{Bierhoff09polymorphicfractional}
\bibinfo{author}{Bierhoff, K.}, \bibinfo{author}{Beckman, N.E.},
  \bibinfo{author}{Aldrich, J.}, \bibinfo{year}{2009}a.
\newblock \bibinfo{title}{{Polymorphic fractional permission inference}}, in:
  \bibinfo{booktitle}{Proceedings of the 18th ACM SIGSOFT International
  Symposium on Software Testing and Analysis}.
\bibitem[{Bierhoff et~al.(2009b)Bierhoff, Beckman and Aldrich}]{BierhoffBA09}
\bibinfo{author}{Bierhoff, K.}, \bibinfo{author}{Beckman, N.E.},
  \bibinfo{author}{Aldrich, J.}, \bibinfo{year}{2009}b.
\newblock \bibinfo{title}{{Practical API Protocol Checking with Access
  Permissions}}, in: \bibinfo{booktitle}{Proceedings of the 23rd European
  Conference on Object-Oriented Programming}, pp. \bibinfo{pages}{195--219}.
\bibitem[{Blom and Huisman(2014)}]{Blom2014ThePrograms}
\bibinfo{author}{Blom, S.}, \bibinfo{author}{Huisman, M.},
  \bibinfo{year}{2014}.
\newblock \bibinfo{title}{The vercors tool for verification of concurrent
  programs}, in: \bibinfo{booktitle}{International Symposium on Formal
  Methods}, \bibinfo{organization}{Springer}. pp. \bibinfo{pages}{127--131}.
\bibitem[{Boyland(2003)}]{boyland2003checking}
\bibinfo{author}{Boyland, J.}, \bibinfo{year}{2003}.
\newblock \bibinfo{title}{{Checking Interference with Fractional Permissions}},
  in: \bibinfo{booktitle}{Proceedings of the 10th International Conference on
  Static Analysis}, \bibinfo{publisher}{Springer-Verlag}. pp.
  \bibinfo{pages}{55--72}.
\bibitem[{Boyland(2013)}]{Boyland2013FractionalPermissions}
\bibinfo{author}{Boyland, J.}, \bibinfo{year}{2013}.
\newblock \bibinfo{title}{{Fractional Permissions}},
  \bibinfo{publisher}{Springer, Berlin, Heidelberg}, pp.
  \bibinfo{pages}{270--288}.
\bibitem[{Boyland et~al.(2014)Boyland, M\"{u}ller, Schwerhoff and
  Summers}]{Boyland2014ConstraintPermissions}
\bibinfo{author}{Boyland, J.T.}, \bibinfo{author}{M\"{u}ller, P.},
  \bibinfo{author}{Schwerhoff, M.}, \bibinfo{author}{Summers, A.J.},
  \bibinfo{year}{2014}.
\newblock \bibinfo{title}{{Constraint Semantics for Abstract Read
  Permissions}}, in: \bibinfo{booktitle}{Proceedings of 16th Workshop on Formal
  Techniques for Java-like Programs - FTfJP'14}, \bibinfo{publisher}{ACM
  Press}. pp. \bibinfo{pages}{1--6}.
\bibitem[{Bull et~al.()Bull, Smith, Westhead, Henty and Davey}]{jomp}
\bibinfo{author}{Bull, J.M.}, \bibinfo{author}{Smith, L.A.},
  \bibinfo{author}{Westhead, M.D.}, \bibinfo{author}{Henty, D.S.},
  \bibinfo{author}{Davey, R.A.}, .
\newblock \bibinfo{title}{A benchmark suite for high performance java}.
\newblock \bibinfo{journal}{Concurrency: Practice and Experience}
  \bibinfo{volume}{12}, \bibinfo{pages}{375--388}.
\newblock \URLprefix
  \url{https://onlinelibrary.wiley.com/doi/abs/10.1002/1096-9128}.
\bibitem[{Cata\~{n}o et~al.(2014)Cata\~{n}o, Ahmed, Siminiceanu and
  Aldrich}]{CatanoASA14}
\bibinfo{author}{Cata\~{n}o, N.}, \bibinfo{author}{Ahmed, I.},
  \bibinfo{author}{Siminiceanu, R.I.}, \bibinfo{author}{Aldrich, J.},
  \bibinfo{year}{2014}.
\newblock \bibinfo{title}{{A case study on the lightweight verification of a
  multi-threaded task server}}.
\newblock \bibinfo{journal}{Sci. Comput. Program.} \bibinfo{volume}{80},
  \bibinfo{pages}{169--187}.
\bibitem[{Cousot and Cousot(1977)}]{Cousot1977}
\bibinfo{author}{Cousot, P.}, \bibinfo{author}{Cousot, R.},
  \bibinfo{year}{1977}.
\newblock \bibinfo{title}{{Abstract interpretation: a unified lattice model for
  static analysis of programs by construction or approximation of fixpoints}},
  in: \bibinfo{booktitle}{Proceedings of the 4th ACM SIGACT-SIGPLAN symposium
  on Principles of programming languages}, \bibinfo{organization}{ACM}. pp.
  \bibinfo{pages}{238--252}.
\bibitem[{Das and Fujimoto(1993)}]{healthsystem}
\bibinfo{author}{Das, S.R.}, \bibinfo{author}{Fujimoto, R.M.},
  \bibinfo{year}{1993}.
\newblock \bibinfo{title}{A performance study of the cancelback protocol for
  time warp}.
\newblock \bibinfo{journal}{SIGSIM Simul. Dig.} \bibinfo{volume}{23},
  \bibinfo{pages}{135--142}.
\newblock \URLprefix \url{http://doi.acm.org/10.1145/174134.158476},
  \DOIprefix\doi{10.1145/174134.158476}.
\bibitem[{Dohrau et~al.(2018)Dohrau, Summers, Urban, M{\"u}nger and
  M{\"u}ller}]{Dohrau2018PermissionPrograms}
\bibinfo{author}{Dohrau, J.}, \bibinfo{author}{Summers, A.J.},
  \bibinfo{author}{Urban, C.}, \bibinfo{author}{M{\"u}nger, S.},
  \bibinfo{author}{M{\"u}ller, P.}, \bibinfo{year}{2018}.
\newblock \bibinfo{title}{Permission inference for array programs}, in:
  \bibinfo{editor}{Chockler, H.}, \bibinfo{editor}{Weissenbacher, G.} (Eds.),
  \bibinfo{booktitle}{Computer Aided Verification},
  \bibinfo{publisher}{Springer International Publishing},
  \bibinfo{address}{Cham}. pp. \bibinfo{pages}{55--74}.
\bibitem[{Duran et~al.(2009)Duran, Teruel, Ferrer, Martorell and Ayguade}]{BOT}
\bibinfo{author}{Duran, A.}, \bibinfo{author}{Teruel, X.},
  \bibinfo{author}{Ferrer, R.}, \bibinfo{author}{Martorell, X.},
  \bibinfo{author}{Ayguade, E.}, \bibinfo{year}{2009}.
\newblock \bibinfo{title}{Barcelona openmp tasks suite: A set of benchmarks
  targeting the exploitation of task parallelism in openmp}, in:
  \bibinfo{booktitle}{Proceedings of the 2009 International Conference on
  Parallel Processing}, \bibinfo{publisher}{IEEE Computer Society},
  \bibinfo{address}{Washington, DC, USA}. pp. \bibinfo{pages}{124--131}.
\newblock \URLprefix \url{https://doi.org/10.1109/ICPP.2009.64},
  \DOIprefix\doi{10.1109/ICPP.2009.64}.
\bibitem[{Ferrara and M\"{u}ller(2012)}]{Ferrara2012}
\bibinfo{author}{Ferrara, P.}, \bibinfo{author}{M\"{u}ller, P.},
  \bibinfo{year}{2012}.
\newblock \bibinfo{title}{{Automatic Inference of Access Permissions}}, in:
  \bibinfo{booktitle}{Verification, Model Checking, and Abstract
  Interpretation: 13th International Conference, VMCAI 2012, Philadelphia, PA,
  USA, January 22-24, 2012. Proceedings}, \bibinfo{publisher}{Springer Berlin
  Heidelberg}. pp. \bibinfo{pages}{202--218}.
\bibitem[{Fonseca et~al.(2016)Fonseca, Cabral, Rafael and
  Correia}]{fonseca2016automatic}
\bibinfo{author}{Fonseca, A.}, \bibinfo{author}{Cabral, B.},
  \bibinfo{author}{Rafael, J.}, \bibinfo{author}{Correia, I.},
  \bibinfo{year}{2016}.
\newblock \bibinfo{title}{{Automatic Parallelization: Executing Sequential
  Programs on a Task-Based Parallel Runtime}}.
\newblock \bibinfo{journal}{International Journal of Parallel Programming}
  \bibinfo{volume}{44}, \bibinfo{pages}{1337--1358}.
\bibitem[{Girard(1987)}]{girard1987linear}
\bibinfo{author}{Girard, J.Y.}, \bibinfo{year}{1987}.
\newblock \bibinfo{title}{{Linear logic}}.
\newblock \bibinfo{journal}{Theoretical computer science} \bibinfo{volume}{50},
  \bibinfo{pages}{1--101}.
\bibitem[{Heule et~al.(2013)Heule, Leino, M\"{u}ller and
  Summers}]{heule2013abstract}
\bibinfo{author}{Heule, S.}, \bibinfo{author}{Leino, K.R.M.},
  \bibinfo{author}{M\"{u}ller, P.}, \bibinfo{author}{Summers, A.J.},
  \bibinfo{year}{2013}.
\newblock \bibinfo{title}{{Abstract read permissions: Fractional permissions
  without the fractions}}, in: \bibinfo{booktitle}{International Workshop on
  Verification, Model Checking, and Abstract Interpretation},
  \bibinfo{organization}{Springer}. pp. \bibinfo{pages}{315--334}.
\bibitem[{Huisman and Mostowski(2015)}]{Huisman2015AReasoning}
\bibinfo{author}{Huisman, M.}, \bibinfo{author}{Mostowski, W.},
  \bibinfo{year}{2015}.
\newblock \bibinfo{title}{{A symbolic approach to permission accounting for
  concurrent reasoning}}, in: \bibinfo{booktitle}{Proceedings - IEEE 14th
  International Symposium on Parallel and Distributed Computing, ISPDC 2015}.
\bibitem[{Huth and Ryan(2000)}]{CTL-logic}
\bibinfo{author}{Huth, M.R.A.}, \bibinfo{author}{Ryan, M.},
  \bibinfo{year}{2000}.
\newblock \bibinfo{title}{Logic in Computer Science: Modelling and Reasoning
  About Systems}.
\newblock \bibinfo{publisher}{Cambridge University Press},
  \bibinfo{address}{New York, NY, USA}.
\bibitem[{Jacobs et~al.(2018)Jacobs, Bosnacki and Kuiper}]{Jacobs:2018}
\bibinfo{author}{Jacobs, B.}, \bibinfo{author}{Bosnacki, D.},
  \bibinfo{author}{Kuiper, R.}, \bibinfo{year}{2018}.
\newblock \bibinfo{title}{Modular termination verification of single-threaded
  and multithreaded programs}.
\newblock \bibinfo{journal}{ACM TOPLAS} \bibinfo{volume}{40}.
\bibitem[{Jacobs et~al.(2011)Jacobs, Smans, Philippaerts, Vogels, Penninckx and
  Piessens}]{jacobs2011verifast}
\bibinfo{author}{Jacobs, B.}, \bibinfo{author}{Smans, J.},
  \bibinfo{author}{Philippaerts, P.}, \bibinfo{author}{Vogels, F.},
  \bibinfo{author}{Penninckx, W.}, \bibinfo{author}{Piessens, F.},
  \bibinfo{year}{2011}.
\newblock \bibinfo{title}{{VeriFast: A powerful, sound, predictable, fast
  verifier for C and java}}, in: \bibinfo{booktitle}{NASA Formal Methods
  Symposium}, \bibinfo{organization}{Springer}. pp. \bibinfo{pages}{41--55}.
\bibitem[{Juhasz et~al.(2014)Juhasz, Kassios, M{\"{u}}ller, Novacek, Schwerhoff
  and Summers}]{Juhasz2014Viper}
\bibinfo{author}{Juhasz, U.}, \bibinfo{author}{Kassios, I.T.},
  \bibinfo{author}{M{\"{u}}ller, P.}, \bibinfo{author}{Novacek, M.},
  \bibinfo{author}{Schwerhoff, M.}, \bibinfo{author}{Summers, A.J.},
  \bibinfo{year}{2014}.
\newblock \bibinfo{title}{{Viper}}.
\newblock \bibinfo{type}{Technical Report}.
\newblock \URLprefix
  \url{https://www.research-collection.ethz.ch/handle/20.500.11850/85086}.
\bibitem[{Le et~al.(2012)Le, Chin and Teo}]{Le2012}
\bibinfo{author}{Le, D.K.}, \bibinfo{author}{Chin, W.N.}, \bibinfo{author}{Teo,
  Y.M.}, \bibinfo{year}{2012}.
\newblock \bibinfo{title}{{Variable Permissions for Concurrency Verification}},
  in: \bibinfo{booktitle}{Formal Methods and Software Engineering: 14th
  International Conference on Formal Engineering Methods, ICFEM 2012, Kyoto,
  Japan, November 12-16, 2012. Proceedings}. \bibinfo{publisher}{Springer
  Berlin Heidelberg}, pp. \bibinfo{pages}{5--21}.
\bibitem[{Lea(2000)}]{JFJF}
\bibinfo{author}{Lea, D.}, \bibinfo{year}{2000}.
\newblock \bibinfo{title}{A java fork/join framework}, in:
  \bibinfo{booktitle}{Proceedings of the ACM 2000 Conference on Java Grande},
  \bibinfo{publisher}{ACM}, \bibinfo{address}{New York, NY, USA}. pp.
  \bibinfo{pages}{36--43}.
\newblock \URLprefix \url{http://doi.acm.org/10.1145/337449.337465},
  \DOIprefix\doi{10.1145/337449.337465}.
\bibitem[{Leavens et~al.(2006)Leavens, Baker and Ruby}]{leavenspreliminary06}
\bibinfo{author}{Leavens, G.T.}, \bibinfo{author}{Baker, A.L.},
  \bibinfo{author}{Ruby, C.}, \bibinfo{year}{2006}.
\newblock \bibinfo{title}{{Preliminary design of JML: A behavioral interface
  specification language for Java}}.
\newblock \bibinfo{journal}{ACM SIGSOFT Software Engineering Notes}
  \bibinfo{volume}{31}, \bibinfo{pages}{1--38}.
\bibitem[{Leino and M\"{u}ller(2009)}]{leino2009basis}
\bibinfo{author}{Leino, K.R.M.}, \bibinfo{author}{M\"{u}ller, P.},
  \bibinfo{year}{2009}.
\newblock \bibinfo{title}{{A basis for verifying multi-threaded programs}}, in:
  \bibinfo{booktitle}{European Symposium on Programming},
  \bibinfo{organization}{Springer}. pp. \bibinfo{pages}{378--393}.
\bibitem[{Leino et~al.(2009)Leino, M\"{u}ller and Smans}]{Chalice2009}
\bibinfo{author}{Leino, K.R.M.}, \bibinfo{author}{M\"{u}ller, P.},
  \bibinfo{author}{Smans, J.}, \bibinfo{year}{2009}.
\newblock \bibinfo{title}{{Verification of Concurrent Programs with Chalice}},
  in: \bibinfo{booktitle}{Foundations of Security Analysis and Design V: FOSAD
  2007/2008/2009 Tutorial Lectures}. \bibinfo{publisher}{Springer Berlin
  Heidelberg}, pp. \bibinfo{pages}{195--222}.
\bibitem[{Meyer(1988)}]{Meyer1988Object-OrientedConstruction}
\bibinfo{author}{Meyer, B.}, \bibinfo{year}{1988}.
\newblock \bibinfo{title}{{Object-Oriented Software Construction}}.
\newblock \bibinfo{edition}{1st} ed., \bibinfo{publisher}{Prentice-Hall, Inc.}
\bibitem[{M\"{u}ller et~al.(2017)M\"{u}ller, Schwerhoff and
  Summers}]{Muller2017Viper:Reasoning}
\bibinfo{author}{M\"{u}ller, P.}, \bibinfo{author}{Schwerhoff, M.},
  \bibinfo{author}{Summers, A.J.}, \bibinfo{year}{2017}.
\newblock \bibinfo{title}{{Viper: A verification infrastructure for
  permission-based reasoning}}, in: \bibinfo{booktitle}{Dependable Software
  Systems Engineering}.
\bibitem[{Odersky et~al.(2004)Odersky, Altherr, Cremet, Emir, Maneth,
  Micheloud, Mihaylov, Schinz, Stenman and Zenger}]{scalaodersky2004}
\bibinfo{author}{Odersky, M.}, \bibinfo{author}{Altherr, P.},
  \bibinfo{author}{Cremet, V.}, \bibinfo{author}{Emir, B.},
  \bibinfo{author}{Maneth, S.}, \bibinfo{author}{Micheloud, S.},
  \bibinfo{author}{Mihaylov, N.}, \bibinfo{author}{Schinz, M.},
  \bibinfo{author}{Stenman, E.}, \bibinfo{author}{Zenger, M.},
  \bibinfo{year}{2004}.
\newblock \bibinfo{title}{{An overview of the Scala programming language}}.
\newblock \bibinfo{type}{Technical Report}. Swiss Federal Institute of
  Technology Lausanne.
\bibitem[{Reynolds(2002)}]{reynolds2002separation}
\bibinfo{author}{Reynolds, J.C.}, \bibinfo{year}{2002}.
\newblock \bibinfo{title}{{Separation logic: A logic for shared mutable data
  structures}}, in: \bibinfo{booktitle}{Logic in Computer Science, 2002.
  Proceedings. 17th Annual IEEE Symposium on}, \bibinfo{organization}{IEEE}.
  pp. \bibinfo{pages}{55--74}.
\bibitem[{Roux and Siminiceanu(2010)}]{RouxS10}
\bibinfo{author}{Roux, P.}, \bibinfo{author}{Siminiceanu, R.},
  \bibinfo{year}{2010}.
\newblock \bibinfo{title}{{Model Checking with Edge-valued Decision Diagrams}},
  in: \bibinfo{booktitle}{Second Formal Methods Symposium, NASA STI, Washington
  D.C., USA, April 13-15, 2010. Proceedings}, pp. \bibinfo{pages}{222--226}.
\bibitem[{Siminiceanu et~al.(2012)Siminiceanu, Ahmed and
  Cata\~{n}o}]{SiminiceanuAC12}
\bibinfo{author}{Siminiceanu, R.I.}, \bibinfo{author}{Ahmed, I.},
  \bibinfo{author}{Cata\~{n}o, N.}, \bibinfo{year}{2012}.
\newblock \bibinfo{title}{{Automated Verification of Specifications with
  Typestates and Access Permissions}}.
\newblock \bibinfo{journal}{ECEASST} \bibinfo{volume}{53}.
\bibitem[{Stork et~al.(2009)Stork, Marques and Aldrich}]{StorkMA09}
\bibinfo{author}{Stork, S.}, \bibinfo{author}{Marques, P.},
  \bibinfo{author}{Aldrich, J.}, \bibinfo{year}{2009}.
\newblock \bibinfo{title}{{Concurrency by default: using permissions to express
  dataflow in stateful programs}}, in: \bibinfo{booktitle}{Proceedings of the
  24th ACM SIGPLAN conference companion on Object oriented programming systems
  languages and applications}, pp. \bibinfo{pages}{933--940}.
\bibitem[{Stork et~al.(2014)Stork, Naden, Sunshine, Mohr, Fonseca, Marques and
  Aldrich}]{Aeminium2014}
\bibinfo{author}{Stork, S.}, \bibinfo{author}{Naden, K.},
  \bibinfo{author}{Sunshine, J.}, \bibinfo{author}{Mohr, M.},
  \bibinfo{author}{Fonseca, A.}, \bibinfo{author}{Marques, P.},
  \bibinfo{author}{Aldrich, J.}, \bibinfo{year}{2014}.
\newblock \bibinfo{title}{{{Æ}minium: A Permission-Based Concurrent-by-Default
  Programming Language Approach}}.
\newblock \bibinfo{journal}{ACM Trans. Program. Lang. Syst.}
  \bibinfo{volume}{36}, \bibinfo{pages}{1--42}.
\bibitem[{Strom and Yemini(1986)}]{strom1986typestate}
\bibinfo{author}{Strom, R.E.}, \bibinfo{author}{Yemini, S.},
  \bibinfo{year}{1986}.
\newblock \bibinfo{title}{{Typestate: A programming language concept for
  enhancing software reliability}}.
\newblock \bibinfo{journal}{IEEE Transactions on Software Engineering} ,
  \bibinfo{pages}{157--171}.
\bibitem[{Sutter and Larus(2005)}]{Sutter}
\bibinfo{author}{Sutter, H.}, \bibinfo{author}{Larus, J.},
  \bibinfo{year}{2005}.
\newblock \bibinfo{title}{{Software and the Concurrency Revolution}}.
\newblock \bibinfo{journal}{Queue} \bibinfo{volume}{3},
  \bibinfo{pages}{54--62}.
\bibitem[{Walter(2016)}]{Walter2016AutomaticInterpretation}
\bibinfo{author}{Walter, S.}, \bibinfo{year}{2016}.
\newblock \bibinfo{title}{{Automatic Inference of Quantified Permissions by
  Abstract Interpretation}}.
\newblock \bibinfo{type}{Technical Report}. Department of Computer Science, ETH
  Zurich.
\newblock \URLprefix
  \url{https://www.ethz.ch/content/dam/ethz/special-interest/infk/chair-program-method/pm/documents/Education/Theses/Seraiah_Walter_MA_report.pdf}.
\bibitem[{Wickerson et~al.(2010)Wickerson, Dodds and
  Parkinson}]{Wickerson2010ExplicitReasoning}
\bibinfo{author}{Wickerson, J.}, \bibinfo{author}{Dodds, M.},
  \bibinfo{author}{Parkinson, M.}, \bibinfo{year}{2010}.
\newblock \bibinfo{title}{Explicit stabilisation for modular rely-guarantee
  reasoning}, in: \bibinfo{editor}{Gordon, A.D.} (Ed.),
  \bibinfo{booktitle}{Programming Languages and Systems},
  \bibinfo{publisher}{Springer Berlin Heidelberg}, \bibinfo{address}{Berlin,
  Heidelberg}. pp. \bibinfo{pages}{610--629}.

\end{thebibliography}
\onecolumn
\newpage
\appendix
\renewcommand{\thesection}{A.\arabic{section}}
\section*{Appendix}

\section{Syntactic Rules for Modelling Object Accesses in a program}\label{rules}

\subsection{Modelling Object Accesses in the Current Method}
This section presents the syntactic rules that capture expression statements in a method and that have been used to support the data-flow and alias-flow analysis of the source code during the metadata extraction phase of the permission inference approach, as explained in Section \ref{meta-analysis}.
The rules are being divided into two main types depending on their access by the current method and other methods: a) \code{Statement} rules, and b) \code{Context} rules, during the analysis. The rules are further categorized according to the type of expressions such as \code{<FieldAccess>}, \code{<Assignment>} or \code{<MethodInvocation>}, encountered during each expression statement and the type of reference variable (\var, \localrefvar, \lvar) accessed in each expression. During parsing, each expression is recursively parsed to fetch the type of expressions designed as based cases. The extracted dependencies are then mapped in the form of a permission-based graph model as explained previously in Section \ref{GC}.

The rules follow the style of sequent calculus in Linear logic, with connectives and implication ($\multimap$) operator, to enforce the strong and constructive interpretation of the specified rules as formulas, to support the analysis.

A complete list of the modelling rules is given below.

\newpage

\begin{enumerate}[label=\Roman*.,leftmargin={0cm}]

\item \textbf{Statement Rules for the Global References}

\[{
\resizebox{\linewidth}{!}{
\begin{tabular}{l l}
\begin{prooftree}
\,\texttt{<Type> \var}\,
\using{\textsf{(\textbf{GR-Decl, \var})}}
\justifies
\begin{array}{l}
\,\texttt{<do-nothing>}\,\\
\end{array}
\end{prooftree}
\\\\\\
\begin{prooftree}
\,\texttt{\var|super.\var|this.\var|<ClassName>.\var|<obj>.\var}\,
\using{\textsf{(\textbf{GR-Read-Only, \var})}}
\justifies
\begin{array}{l}
\,\texttt{addReadEdge(\foo, \var)}\, \\
\end{array}
\end{prooftree}
\\\\\\
\begin{prooftree}
\,\texttt{[PRIM\_TYPE] \var = \varnum{1}|<LITERAL>}\,
\using{\textsf{(\textbf{GR-Val-Flow,\var})}}
\justifies
\begin{array}{l}
\,\texttt{addWriteEdge(\foo, \var)($\forall a \in$ aliasOf(\var) addWriteEdge(\foo, a)),(apply(GR-Read-Only, \varnum{1})}\,\\\\
\end{array}
\end{prooftree}
\\\\\\
\begin{prooftree}
\,\texttt{[<Type>] \var = new <Type>(\varnum{2})|<Number\_Literal>}\,
\using{\textsf{(\textbf{GR-New-Obj, \var})}}
\justifies
\begin{array}{l}
\,\texttt{(addWriteEdge(\foo,\var) $\multimap$ apply(Context-N,\var)),(apply(GR-Read-Only, \varnum{2})}\,\\
\end{array}
\end{prooftree}
\\\\\\
\begin{prooftree}
\,\texttt{[<REF\_TYPE>] \var = \varnum{1}}\,
\using{\textsf{(\textbf{GR-Add-Flow, \var})}}
\justifies
\begin{array}{l}
\,\texttt{($\exists$aliasEdge(\var, \varnum{2}) $\multimap$ removeAliasEdge(\var, \varnum{2})), addAliasEdge(\var, \varnum{1})}\,\\
\,\texttt{apply(GR-Read-Only, \varnum{1})}\,\\
\end{array}
\end{prooftree}
\\\\\\
\begin{prooftree}
\,\texttt{[<REF\_TYPE>] \var = \localrefvar}\,
\using{\textsf{(\textbf{GR-Addr-Flow, \localrefvar})}}
\justifies
\begin{array}{l}
\,\texttt{($\exists$aliasEdge(\localrefvar, \varnum{1})$\multimap$addAliasEdge(\var, \localrefvar))}\, \\\\
\,\texttt{($\exists$aliasEdge(\var, \varnum{2})$\multimap$removeAliasEdge(\var, \varnum{2})),apply(GR-Read-Only, \varnum{1})}\,
\end{array}
\end{prooftree}
\end{tabular}}
}
\]
{\small
\[{
\resizebox{\linewidth}{!}{
\begin{tabular}{l l}
\begin{prooftree}
\,\texttt{[<REF\_TYPE>] \var = \lvar}\,
\using{\textsf{(\textbf{GR-Addr-Flow, \lvar}})}
\justifies
\begin{array}{l}
\,\texttt{($\exists$aliasEdge(\var, \varnum{1})$\multimap$removeAliasEdge(\var, \varnum{1}))}\,\\
\end{array}
\end{prooftree}
\\\\\\
\begin{prooftree}
\,\texttt{<Type> \var = <Null\_Literal>|MCall(<post-perm>,\varnum{1})}\,
\using{\textsf{(\textbf{GR-NullAddr-Init, \var})}}
\justifies
\begin{array}{l}
\,\texttt{<do-nothing>|MCall(<post-perm>,\varnum{1})}\,
\end{array}
\end{prooftree}
\\\\\\
\begin{prooftree}
\,\texttt{\var = <Null\_Literal>|MCall(<post-perm>,\varnum{2})}\,
\using{\textsf{{(\textbf{GR-NullAddr-Flow, \var})}}}
\justifies
\begin{array}{l}
\,\texttt{addWriteEdge(\foo, \var)($\exists$aliasEdge(\var, \varnum{1})$\multimap$removeAliasEdge(\var, \varnum{1})}\, \\
\,\texttt{apply(Context-N, \var)),($\forall$ a $\in$ aliasOf(\var),apply(<GR-NullAddr-Flow>, a)))\,}\\
\,\texttt{,apply(MCall(<post-perm>, \varnum{1}))}\\
\end{array}
\end{prooftree}
\\\\\\
\begin{prooftree}
\,\texttt{\var = \localrefvar}\,
\using{\textsf{(\textbf{GR-SelfAddr-Flow, \localrefvar})}}
\justifies
\begin{array}{l}
\,\texttt{($\exists$aliasEdge(\localrefvar, \var)$\multimap$<do-nothing>)}\,\\
\end{array}
\end{prooftree}
\\\\\\
\begin{prooftree}
\,\texttt{\var = \var}\,
\using{\textsf{(\textbf{GR-SelfAddr-Flow, \var})}}
\justifies
\begin{array}{l}
\,\texttt{<do-nothing>}\,\\
\end{array}
\end{prooftree}
\end{tabular}}
}
\]}
\\
\item \textbf{Statement Rules for the Local References}\label{LR}
{\small
\[{
\resizebox{\linewidth}{!}{
\begin{tabular}{l l}
\begin{prooftree}
\,\texttt{\localrefvar}\,
\using{\textsf{(\textbf{LR-Read-Only, \localrefvar})}}
\justifies
\begin{array}{l}
\,\texttt{($\exists$aliasEdge(\localrefvar, \var)$\multimap$(apply(GR-Read-Only(\var))}\, \\
\end{array}
\end{prooftree}
\\\\\\
\begin{prooftree}
\,\texttt{\localrefvar.\varnum{1} = <Number\_Literal>}\,
\using{\textsf{(\textbf{LR-Val-Flow, \localrefvar})}}
\justifies
\begin{array}{l}
\,\texttt{($\exists$aliasEdge(\localrefvar, \var)$\multimap$apply(GR-Val-Flow, \varnum{1}))}\, \\
\end{array}
\end{prooftree}
\\\\\\
\begin{prooftree}
\,\texttt{[<REF\_TYPE>] \localrefvar = \var}\,
\using{\textsf{{(\textbf{LR-Addr-Flow, \var})}}}
\justifies
\begin{array}{l}
\,\texttt{($\exists$aliasEdge(\localrefvar, \varnum{1})$\multimap$removeAliasEdge(\localrefvar,\varnum{1})),addAliasEdge(\localrefvar,\var))}\, \\\\
\,\texttt{apply(GR-Read-Only, \var)}\,
\end{array}
\end{prooftree}
\\\\\\
\begin{prooftree}
\,\texttt{[<REF\_TYPE>]  \localrefvar = \localrefvarnum{1}}\,
\using{\textsf{(\textbf{LR-Addr-Flow, \localrefvar}})}
\justifies
\begin{array}{l}
\,\texttt{($\exists$aliasEdge(\localrefvar,\var)$\multimap$removeAliasEdge(\localrefvar,\var)),}\,\\\\
\,\texttt{($\exists$aliasEdge(\localrefvarnum{1},\varnum{1})$\multimap$addAliasEdge(\localrefvar,\localrefvarnum{1}),apply(GR-Read-Only, \varnum{1}))}\,
\end{array}
\end{prooftree}
\\\\\\
\begin{prooftree}
\,\texttt{[<REF\_TYPE>] \localrefvar = \lvar}\,
\using{\textsf{(\textbf{LR-Addr-Flow, \lvar})}}
\justifies
\begin{array}{l}
\,\texttt{($\exists$aliasEdge(\localrefvar, \var)$\multimap$removeAliasEdge(\localrefvar,\var))}\, \\
\end{array}
\end{prooftree}
\\\\\\
\begin{prooftree}
\,\texttt{\localrefvar = new <[<REF\_TYPE>]>(\varnum{2}|<Number\_Literal>}\,
\using{\textsf{(\textbf{LR-New-Obj, \localrefvar})}}
\justifies
\begin{array}{l}
\,\texttt{($\exists$aliasEdge(\localrefvar,\varnum{1})$\multimap$removeAliasEdge(\localrefvar,\varnum{1})}\,\\\\
\,\texttt{(apply(GR-Read-Only,\varnum{2}))}\,
\end{array}
\end{prooftree}
\\\\\\
\begin{prooftree}
\,\texttt{\localrefvar = <Null\_Literal>|MCall(<post-perm>,\var)}\,
\using{\textsf{(\textbf{LR-NullAddr-Flow, \localrefvar})}}
\justifies
\begin{array}{l}
\,\texttt{($\exists$aliasEdge(\localrefvar, \var)$\multimap$removeAliasEdge(\localrefvar, \var),}\, \\\\
\,\texttt{($\forall$ a $\in$ aliasOf(\localrefvar),apply(<GR-NullAddr-Flow>, a)),apply(MCall(<post-perm>,\var))\,
}\end{array}
\end{prooftree}
\\\\\\
\begin{prooftree}
\,\texttt{<Type> \localrefvar = <Null\_Literal>|MCall(<post-perm>,\var)}\,
\using{\textsf{(\textbf{LR-NullAddr-Init, \localrefvar})}}
\justifies
\begin{array}{l}
\,\texttt{<do-nothing>)|MCall(<post-perm>,\var)}\,
\end{array}
\end{prooftree}
\\\\\\
\begin{prooftree}
\,\texttt{\localrefvar = \var}\,
\using{\textsf{(\textbf{LR-SelfAddr-Flow, \localrefvar})}}
\justifies
\begin{array}{l}
\,\texttt{($\exists$aliasEdge(\localrefvar,\var) $\multimap$ <do-nothing>}\,\\
\end{array}
\end{prooftree}
\\\\\\
\begin{prooftree}
\,\texttt{\localrefvarnum{1} = \localrefvarnum{1}}\,
\using{\textsf{(\textbf{LR-SelfAddr-Flow, \localrefvar})}}
\justifies
\begin{array}{l}
\,\texttt{<do-nothing>}\,\\\\
\end{array}
\end{prooftree}
\end{tabular}}
}
\]}

\item \textbf{Statement Rules for Method Calls}\label{MC}

\end{enumerate}


{\scriptsize
\[{
\begin{tabular}{l l}
\begin{prooftree}
\,\texttt{MCall([<args>])|super.MCall([<args>])|super([<args>])}\,
\using{\textsf{\textbf{MCall(Immutable, \var)}}}
\justifies
\begin{array}{l}
\,\texttt{addReadEdge(\var, \foo) apply(Context-R, \var)}\, \\
\end{array}
\end{prooftree}
\\\\
\begin{prooftree}
\,\texttt{MCall([<args>])|super.MCall([<args>])|super([<args>])}\,
\using{\textsf{\textbf{MCall(Pure, \var)}}}
\justifies
\begin{array}{l}
\,\texttt{addReadEdge(\foo, \var), apply(Context-RW, \var)}\, \\
\end{array}
\end{prooftree}
\\\\
\begin{prooftree}
\,\texttt{MCall([<args>])|super.MCall([<args>])|super([<args>])}\,
\using{\textsf{\textbf{MCall(Full, \var)}}}
\justifies
\begin{array}{l}
\,\texttt{addWriteEdge(\foo, \var), apply(Context-R, \var)}\, \\
\end{array}
\end{prooftree}
\\\\
\begin{prooftree}
\,\texttt{MCall([<args>])|super.MCall([<args>])|super([<args>])}\,
\using{\textsf{\textbf{MCall(Share, \var)}}}
\justifies
\begin{array}{l}
\,\texttt{addWriteEdge(\foo, \var) apply(Context-RW, \var)}\, \\
\end{array}
\end{prooftree}
\\\\
\begin{prooftree}
\,\texttt{MCall([<args>])|super.MCall([<args>])|super([<args>])}\,
\using{\textsf{\textbf{MCall(Unique, \var)}}}
\justifies
\begin{array}{l}
\,\texttt{addWriteEdge(\foo, \var), apply(Context-N, \var)}\, \\
\end{array}
\end{prooftree}
\\\\
\begin{prooftree}
\,\texttt{MCall([<args>])|super.MCall([<args>])|super([<args>])}\,
\using{\textsf{\textbf{MCall(None, \var)}}}
\justifies
\begin{array}{l}
\,\texttt{<do-nothing>}\,\\
\end{array}
\end{prooftree}
\end{tabular}
}\]}

\subsection{Modelling Object Accesses through Other Methods}\label{context-rules}

As discussed previously in Section \ref{GC}, the context rules model the read, write behavior of other methods on the shared objects accessed in the current method.\\

\begin{enumerate}[label=\Roman*.]
\setlength{\itemsep}{6pt}
\item \textbf{Context Rules for the Global References}
{
\footnotesize
\[{
\resizebox{\linewidth}{!}{
\begin{tabular}{l l}
\begin{prooftree}
\,\texttt{\var}\,
\using{\textsf{\textbf{(Context-R, \var)}}}
\justifies
\begin{array}{l}
\,\texttt{addReadEdge(context, \var)}\, \\
\end{array}
\end{prooftree}
\\\\
\begin{prooftree}
\,\texttt{\var}\,
\using{\textsf{\textbf{(Context-RW, \var)}}}
\justifies
\begin{array}{l}
\,\texttt{addReadEdge(context, \var), addWriteEdge(context, \var)}\, \\\\
\end{array}
\end{prooftree}
\\\\
\begin{prooftree}
\,\texttt{\var}\,
\using{\textsf{\textbf{(Context-N, \var)}}}
\justifies
\begin{array}{l}
\,\texttt{removeReadEdge(\context, \var), removeWriteEdge(\context, \var)}\, \\
\end{array}
\end{prooftree}
\end{tabular}}}
\]}

\hspace{0.75cm}\item \textbf{Context Rules for the Local References}
{
\footnotesize
\[{
\resizebox{\linewidth}{!}{
\begin{tabular}{l l}
\begin{prooftree}
\,\texttt{\localrefvar}\,
\using{\textsf{\textbf{(Context-R, \var)}}}
\justifies
\begin{array}{l}
\,\texttt{($\exists$aliasEdge(\localrefvar, \var) $\multimap$ apply(Context-R, \var))}\, \\
\end{array}
\end{prooftree}
\\\\
\begin{prooftree}
\,\texttt{\localrefvar}\,
\using{\textsf{\textbf{(Context-RW, \var)}}}
\justifies
\begin{array}{l}
\,\texttt{($\exists$aliasEdge(\localrefvar, \var) $\multimap$ apply(Context-RW, \var))}\, \\\\
\end{array}
\end{prooftree}
\\\\
\begin{prooftree}
\,\texttt{\localrefvar}\,
\using{\textsf{\textbf{(Context-N, \var)}}}
\justifies
\begin{array}{l}
\,\texttt{($\exists$aliasEdge(\localrefvar, \var) $\multimap$ apply(Context-N, \var))}\, \\\\
\end{array}
\end{prooftree}
\end{tabular}}}
\]}
\end{enumerate}

\section{Access Permission Inference Rules}\label{AP-rules}
This section lists the access permission inference rules used to generate five types of symbolic permissions on the objects referenced in a method, as explained previously in Section \ref{GT}.
{\normalsize
\[{
\resizebox{\linewidth}{!}{
\begin{tabular}{l l}
\begin{prooftree}
\,\texttt{$\neg$$\exists$readEdge(context,\var) $\wedge$ $\neg$$\exists$writeEdge(context, \var)$\exists$readEdge(\foo, \var) $\wedge$ $\exists$writeEdge(\foo, \var)}\,
\using{\textsf{(\textbf{Unique})}}
\justifies
\begin{array}{l}
\,\texttt{unique(\var)}\,
\end{array}
\end{prooftree}
\\\\\\
\begin{prooftree}
\,\texttt{$\neg$$\exists$writeEdge(\foo,\var) $\wedge$ $\exists$readEdge(\var,\foo) $\wedge$ $\neg$$\exists$writeEdge(context,\var) $\wedge$ $\exists$readEdge(\var,context)}\,
\using{\textsf{(\textbf{Immutable})}}
\justifies
\begin{array}{l}
\,\texttt{immutable(\var)} \,
\end{array}
\end{prooftree}
\\\\\\
\begin{prooftree}
\,\texttt{$\exists$readWriteEdge(\foo,\var) $\wedge$ $\neg$$\exists$writeEdge(context,\var) $\wedge$ $\exists$readEdge(\var,context)}\,
\using{\textsf{(\textbf{Full})}}
\justifies
\begin{array}{l}
\,\texttt{full(\var)} \,
\end{array}
\end{prooftree}
\\\\\\
\bigskip
\begin{prooftree}
\,\texttt{$\exists$writeEdge(\foo,\var) $\wedge$ $\exists$readEdge(\var,\foo) $\wedge$ $\exists$writeEdge(context,\var) $\wedge$ $\exists$readEdge(\var, context)}\,
\using{\textsf{(\textbf{Share})}}
\justifies
\begin{array}{l}
\,\texttt{share(\var)} \,
\end{array}
\end{prooftree}
\\\\\\
\bigskip
\begin{prooftree}
\,\texttt{$\neg$$\exists$writeEdge(\foo,\var) $\wedge$ $\exists$readEdge(\var,context) $\wedge$ $\exists$writeEdge(context,\var)}\,
\using{\textsf{(\textbf{Pure})}}
\justifies
\begin{array}{l}
\,\texttt{pure(\var)}\,
\end{array}
\end{prooftree}
\\\\\\
\bigskip
\begin{prooftree}
\,\texttt{$\neg\exists$writeEdge(\foo,\var) $\wedge \exists$readEdge(\var,\foo) $\wedge \neg\exists$writeEdge(context,\var) $\wedge \neg\exists$readEdge(\var,context)}\,
\using{\textsf{(\textbf{None})}}
\justifies
\begin{array}{l}
\,\texttt{none(\var)}\,
\end{array}
\end{prooftree}
\end{tabular}}
}
\]}

\end{document}